\documentclass[final,3p,times]{elsarticle}

\usepackage[utf8]{inputenc}
\usepackage[T1]{fontenc}
\usepackage{lmodern,textcomp} 

\graphicspath{{graphics/}}

\DeclareGraphicsExtensions{.pdf,.jpeg,.png}

\usepackage{epsfig}
\usepackage{lipsum}

\usepackage{amsmath}
\usepackage{amsfonts}
\usepackage{amssymb}
\usepackage{float}
\usepackage[hyphens]{url}

\usepackage[normalem]{ulem}

\usepackage{booktabs}
\usepackage{tabularx}
\usepackage{threeparttable}
\usepackage{siunitx}
\usepackage[version=4]{mhchem}

\usepackage[colorlinks=true, citecolor=blue, linkcolor=blue, filecolor=blue,urlcolor=blue,breaklinks=true, bookmarks=false]{hyperref}
\usepackage{enumitem}

\usepackage[gen]{eurosym}
\usepackage{multirow}
\usepackage{tikz}
\tikzset{every picture/.style={/utils/exec={\sffamily}}}

\usepackage{tikz}

\usepackage{adjustbox}

\usepackage{fixltx2e}

\journal{Applied Energy}

\begin{document}
\begin{frontmatter}



\title{Are biofuel mandates cost-effective? - an analysis of transport fuels and biomass usage to achieve emissions targets in the European energy system}

\author[chalmers]{M.~Millinger\corref{cor1}}
\ead{markus.millinger@chalmers.se}
\author[chalmers]{L.~Reichenberg}
\author[chalmers]{F.~Hedenus}
\author[chalmers]{G.~Berndes}
\author[tuberlin]{E.~Zeyen}
\author[tuberlin]{T.~Brown}

\cortext[cor1]{Corresponding author}
\address[chalmers]{Department of Space Earth and Environment, Chalmers University of Technology, 412 96, Göteborg, Sweden}
\address[tuberlin]{Department of Digital Transformation in Energy Systems, Technische Universität Berlin, Einsteinufer 25 (TA 8), 10587 Berlin, Germany}

\begin{abstract}
Abatement options for the hard-to-electrify parts of the transport sector are needed to achieve ambitious emissions targets. Biofuels based on biomass, electrofuels based on renewable hydrogen and a carbon source, as well as fossil fuels compensated by carbon dioxide removal (CDR) are the main options. Currently, biofuels are the only renewable fuels available at scale and are stimulated by blending mandates. Here, we estimate the system cost of enforcing such mandates in addition to an overall emissions cap for all energy sectors. We model overnight scenarios for 2040 and 2060 with the sector-coupled European energy system model PyPSA-Eur-Sec, with a high temporal resolution. The following cost drivers are identified: (i) high biomass costs due to scarcity, (ii) opportunity costs for competing usages of biomass for industry heat and combined heat and power (CHP) with carbon capture, and (iii) lower scalability and generally higher cost for biofuels compared to electrofuels and fossil fuels combined with CDR. With a -80\% emissions reduction target in 2040, variable renewables, partial electrification of heat, industry and transport and biomass use for CHP and industrial heat are important for achieving the target at minimal cost, while an abatement of remaining liquid fossil fuel use increases system cost. In this case, a 50\% biofuel mandate increases total energy system costs by 128-229 billion~\texteuro, corresponding to 39-82\% of the liquid fuel cost without a mandate. With a negative -105\% emissions target in 2060, fuel abatement options are necessary, and electrofuels or the use of CDR to offset fossil fuel emissions are both more competitive than biofuels. In this case, a 50\% biofuel mandate increases total costs by 16-31 billion~\texteuro, or 8-14\% of the liquid fuel cost without a mandate. Biomass is preferred in CHP and industry heat, combined with carbon capture to serve negative emissions or electrofuel production, thereby utilising biogenic carbon several times. Sensitivity analyses reveal significant uncertainties but consistently support that higher biofuel mandates lead to higher costs.
\end{abstract}


\end{frontmatter}

\section*{Highlights}

\begin{itemize}[noitemsep]
  \item Biofuel mandates of 50\% increase fuel cost by 39-82\% in the medium term
  \item Electrofuels more cost-effective at negative emissions targets
  \item Biomass use in industry heat and combined heat and power cost-effective
  \item Biomass scarcity, opportunity costs and scalability of other options key factors
\end{itemize}

\section{Introduction}
The transport sector accounted for 30\% of total greenhouse gas (GHG) emissions in the EU-27 in 2019, with an increasing trend \cite{EEA2022}. Recent developments of electric vehicles and targets for phasing out internal combustion engine vehicles (ICEVs) \cite{Wappelhorst2020,EuropeanParliament2021} indicate a considerable electrification of land-based transport within the next few decades. However, a phase-out of ICEVs takes time and thus a liquid hydrocarbon fuel demand persists for several decades even at high electrification rates \cite{Morfeldt2021,Millinger2019}. In maritime transport and aviation, a demand for liquid hydrocarbon fuels is likely to remain also in the long run \cite{IPCC2018,Gray2021,ICCT2020}. Alternative fuel solutions are thus required to achieve ambitious emissions targets. Biofuels and electrofuels are the two available renewable liquid hydrocarbon fuel options \cite{Hannula2019}. Another option is the continued use of unabated fossil fuels combined with carbon dioxide removal (CDR) elsewhere in the system \cite{Fuss2014,Azar2013,Lehtveer2019}.

Currently, conventional biofuels produced from food crops are the dominating option \cite{EuropeanCommission2020}, but they are connected to land use change issues and other sustainability risks \cite{Creutzig2015,Creutzig2016}, and are being phased out in the EU \cite{EuropeanParliament2018}. Biomass residues which are suitable for existing conventional biofuel processes (such as used cooking oil) are scarce \cite{VanGrinsven2020}. Instead, advanced biofuels based on lignocellulosic biomass residues show a relatively large albeit uncertain potential \cite{Ruiz2019}; however, no commercially operational production exists today.

Electrofuels are produced with hydrogen and carbon as feedstocks. Hydrogen can be sourced from electrolysers, which can use electricity when it is cheap in a system dominated by variable renewable energy (VRE). However, the potential depends on a substantial expansion of VRE (or other carbon emissions free) capacity, and thus can be seen as a large-scale option only in the longer term, especially as there is competition for hydrogen from other usages \cite{Millinger2021,Ueckerdt2021}.

An emissions cap-and-trade or tax is often seen as the first-best policy option for achieving targets, since it in theory leads to the least-cost attainment of emissions targets \cite{Jaffe2005}. However, fuel mandates may be important tools for a country or for the EU in a second-best setting if the ideal first-best policy mix is hard to implement or if there are market barriers hindering abatement solutions \cite{Lapan2012,Jaffe2005,Lehmann2013}. Also, mandates may (i) support the development of promising technologies \cite{Jaffe2005}, (ii) focus mitigation efforts to the transport sector which is not pressured by international competition \cite{Hoel1996}, and (iii) count towards other goals, such as improved energy security \cite{Berndes2007}.

In the EU, transport fuels are subject to fuel taxes and blending mandates for achieving emissions targets, and there is a proposal to include the transport and additional industry sectors in the EU-ETS \cite{EuropeanCommission2021}. The proposal for the new Renewable Energy Directive (RED III) \cite{EC2021a} includes renewable fuel mandates for the aviation (20\% in 2035, with at least 5\% electrofuels) \cite{EC2021b} and road transport sectors (2.3\% advanced biofuels for light transport in 2035), while the maritime sector is to reduce its energy intensity by 20\% in 2035 compared to 2020 \cite{EC2021c}. However, several countries have set targets that significantly surpass those of the EU as a whole. For instance, Sweden targets to decrease fuel emissions by 66\% for diesel and 23\% for gasoline by 2030, through blending in biofuels \cite{Regeringskansliet2021}, and Finland aims for 30\% biofuels in the fuel mix by 2030 \cite{BusinessFinland2019}. In the US, the Renewable Fuel Standard currently mandates a blending of around 10\% biofuels into the fuel mix \cite{USEnergyInformationAdministration2021}.

Biofuels present the main short-term option to fulfill fuel mandates \cite{Meisel2020}, by blending them into the fuels used for aviation, road and maritime transport. The fuel mandates thus incentivise investments in biofuel production (supply chain and biorefineries) on a scale to satisfy a sizeable part of the demand for renewable fuels. Although scaling up of new options such as CDR or renewable fuels based on VRE may prove to be challenging \cite{Bento2016,Cherp2021}, the future may also see a large cost reduction for electrolysis \cite{IRENA2020a,IEA2019}, electrofuels \cite{Brynolf2018} and CDR \cite{Keith2018}. In addition, sustainability constraints and competition for biomass may increase biomass prices and thus affect the competitiveness of biofuels as well as of biomass-based CDR options.

A holistic assessment of biomass usage competitions in the energy system requires the inclusion of all energy sectors, which is usually covered in Integrated Assessment Models (IAMs) \cite{IPCC2018,Leblanc2022,Bauer2020,Ahlgren2017} and older energy system optimisation models (ESOMs) such as TIMES/MARKAL \cite{Blanco2018}, with biofuels often emerging as a key solution if CDR is restricted. However, these models generally lack a high spatio-temporal resolution \cite{Iwanaga2021}, which is needed for an explicit representation of VRE \cite{Collins2017,Reichenberg2018, Horsch2017,Nahmmacher2016} and variable production such as electrolysers (and thus electrofuels, which have mostly been lacking as a technology option in such studies). Modern ESOMs employ a higher temporal resolution and have recently been enhanced to encompass all energy sectors (sector-coupled) and a high spatio-temporal resolution for both supply and flexible demand \cite{Victoria2022,Bogdanov2021,Pickering2022}, which enables a holistic analysis of biomass usage and of abatement alternatives for the transport sectors. None of the papers based on these sector-coupled models have specifically targeted biomass use or biofuels, and - to our knowledge - neither have biofuel mandates been investigated in energy system modeling or IAM studies. In order to do this, we expand the sector-coupled European ESOM PyPSA-Eur-Sec \cite{PyPSA-Eur-Sec2021} with details on biomass and bioenergy options.

In this work, we investigate the competition for fuel supply under CO$_2$ emission reduction targets, and the effects of biofuel mandates on energy system costs. We do this by quantifying the increase in total energy system costs as well as liquid fuel costs that biofuel mandates would lead to in the medium (\textasciitilde 2040) and long-term (\textasciitilde2060) perspectives.

\section{Materials and Methods}
We use PyPSA-Eur-Sec \cite{PyPSA-Eur-Sec2021}, which minimises total system costs (including the liquid fuel costs), and analyse system effects and costs when increasing the required minimum share of biofuels that is blended into the liquid hydrocarbon fuel demand. We thus investigate the additional cost of moving away from the optimal use of biomass \cite{Neumann2021} to scenarios which are constrained in terms of dedicated use of biomass for liquid fuel production. These latter scenarios may be viewed as proxies for policies which promote biofuels, i.e.\ biofuel mandates. We investigate this question for the medium (\textasciitilde 2040) and long term (\textasciitilde 2060). The two time horizons are different in terms of CO$_2$ cap as well as other parameters, such as degree of electrification and technology maturity and costs. We assess the effect of carbon sequestration availability, and investigate two different scenarios for domestic biomass potential: one conservative and one more optimistic. In addition, we allow import of biomass to Europe, but at a relatively high cost, as outlined below.

\subsection{Model: PyPSA-Eur-Sec}
PyPSA-Eur-Sec \cite{PyPSA-Eur-Sec2021,Victoria2022} is a sector-coupled full European energy system model including the power sector, transport, space and water heating, industry and industrial feedstocks. The model co-optimizes capacity expansion of energy generation and conversion, as well as their production.

In this work, we expand the model by a rich biomass resource and bioenergy technology portfolio. The further developed version of the model used in this work is available for free use under an open-source license \cite{MillingerPyPSA2021}.

We use a 37-node spatial resolution and an uninterrupted 1-hourly temporal resolution for a full year in overnight brown-field scenarios. The transmission grid is adapted to be a HVDC lossy transport model, and transmission is constrained to increase by max. 50\% in terms of total line volume compared to today.

End-energy demands for the different sectors are calculated based on the JRC IDEES database \cite{Mantzos2017} with additions for non-EU countries \cite{Zeyen2021,Victoria2022}, while energy carrier production including hydrogen, methane and liquid fuels is determined endogenously. Technology data is elaborated on in the supplementary information.

The model runs were performed on the Chalmers Centre for Computational Science and Engineering (C3SE) computing cluster, using 64 threads and 768 GB RAM (or 96 GB RAM for the lower resolution sensitivity runs). The model set-up used here requires up to \textasciitilde400 GB RAM (which can be lowered by using fewer threads).

\subsection{Biomass and Bioenergy}\label{sec:biomass}
\begin{figure}
    \centering
    \includegraphics[trim=0 40 190 78,clip,width=0.8\textwidth]{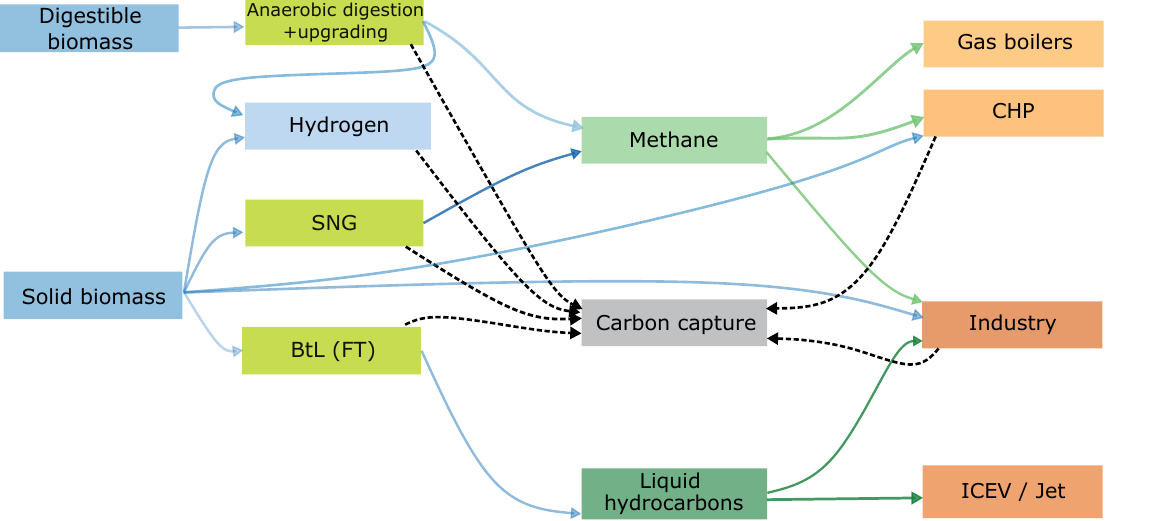}
    \caption{Simplified depiction of the biomass usage options in the model. Energy flows are shown, except for the dashed lines going to carbon capture (which is optional for each of the shown processes), which show mass flows of carbon. The captured carbon can be utilized for hydrocarbon production, or sequestered. Hydrogen can also be produced through electrolysis and steam methane reforming (SMR), and can be used for numerous applications, including FCEVs, electrofuel production and as industry feedstock (not shown.)}
    \label{fig:biomass_pathways}
\end{figure}

A variety of biomass categories and conversion technologies are introduced in the model. Different biomass residue types are clustered into the categories solid biomass and biogas from digestible biomass (Table~\ref{tab:biomassPot}). Solid biomass can be used for a variety of applications in heat, power and fuel production, and can be combined with carbon capture (Figure~\ref{fig:biomass_pathways}). Digestible biomass can be used for biogas production, which is upgraded to biomethane and can likewise be combined with carbon capture.

In the main scenarios, the domestic biomass availability is varied as stated in Table \ref{tab:biomassPot}. Depending on the biomass scenario, solid biomass and biogas from digestible biomass can together provide corresponding to either about 5 or 23\% of the resulting total primary energy demand of around 16 PWh. A weighted average of country-level biomass costs from the \textit{high} biomass scenario for 2050 from the JRC ENSPRESO data base \cite{Ruiz2019} is used and is held constant across scenarios in this study.

\begin{table}[]
    \centering
    \begin{footnotesize}
    \begin{tabular}{l c c c}
    \toprule
                                    & Low  & High & Cost\\
                                    & TWh     & TWh  & \texteuro/MWh\\
                                    \midrule 
        Forest residues             & 267     & 1654 & 12\\
        Industry wood residues      & 76      & 381  & 6\\
        Landscape care              & 42      & 214  & 8\\
        \cmidrule{2-3} \relax
        $\Sigma$ \ Solid biomass   & 385     & 2249 &\\
        \cmidrule{2-3}
        Manure \& slurry            & 173     & 522  & 20\\
        Municipal biowaste          & 122     & 222  & 0.14\\
        Sewage sludge               & 8       & 15   & 17\\
        Straw                       & 186     & 601  & 10\\
        \cmidrule{2-3} \relax
        $\Sigma$ \ Biogas (dig. biomass)  & 489     & 1359 &\\
        \cmidrule{2-3} \relax
        $\Sigma$ \ \textbf{Biomass} & 874 & 3608 &\\
        \bottomrule
    \end{tabular}
    \caption{Domestic biomass scenarios (TWh). The digestible biomass is given in the biogas potential. Values stem from the medium (here denoted \textit{low}) and high biomass potential scenarios from the JRC ENSPRESO data base \cite{Ruiz2019}. A weighted average of country-level biomass costs used from the \textit{high} biomass scenario for 2050 is held constant across scenarios in this study.}
    \label{tab:biomassPot}
    \end{footnotesize}
\end{table}

Only biomass residues and wastes are included in the analysis, i.e.\ bioenergy crops are excluded. The current political and policy context hints towards a limited role for dedicated cultivation in EU, due to concerns about competition with food and risks for environmental impacts from cropland expansion. Thereby, the only option considered for producing liquid biofuels is based on solid biomass (i.e.\ biomass to liquid, BtL), and biofuel imports are excluded. The use of residues and waste as bioenergy feedstock is assumed not to influence the net flow of CO2 between the atmosphere and the biosphere, which is driven by photosynthesis, respiration, decay, and combustion of organic matter. See Section \ref{sec:discussionBiomass} for further elaboration on this carbon neutrality assumption.

\subsubsection{Biomass imports}
Biomass supply and demand in IAMs depend on many different factors, which makes it difficult to construct a global biomass supply curve based on their results. Still, global trade of biomass needs to somehow be represented in a regional ESOM to be more realistic. We use the model comparison by Bauer et al.\ \cite{Bauer2020} which focuses on biomass use in carbon mitigation scenarios, and select five models which represent the competition between biomass supply and food, pasture, and nature, and provide global biomass prices (two of the models also include competition for land for afforestation). For 2050, the global supply varies between 130 and 250 EJ in the different model results, and the biomass price spans between 10 and 21 USD/GJ. Using the average of these models we assume that 175 EJ of biomass can be supplied globally and annually at a price of 15 USD/GJ. We use regional data on biomass use per capita and population estimates \cite{KC2017} to find that 20 EJ biomass may be imported to Europe at a price of 15 \texteuro/GJ. For each additional EJ to be imported the price is assumed to increase by 0.25 \texteuro/GJ, based on the slope of the low-cost scenarios.

In 2021 wood chip prices were at around 8 \texteuro/GJ (30 \texteuro/MWh), i.e.\ the above prices assume a substantial price increase compared to today, which reflects an increased demand for biomass in scenarios complying with stringent GHG emission targets. We test the effect of this assumption on results in a sensitivity analysis. Only solid biomass can be imported, and the import prices are held constant across all scenarios. Direct biofuel imports from outside Europe are excluded.



\subsection{Sector-specific assumptions}\label{sec:transport}
For aviation, increases in traveled passenger kilometers of 50\% by 2040 and 100\% by 2060 in Europe are assumed, compared to 2019 levels. Efficiency improvements of 3\% and 20\% in 2040 and 2060, respectively, are assumed \cite[][extrapolated for 2060]{ATAG2020}. Based on this, we assume a fuel demand increase compared to 2019 of 50\% and 70\% in 2040 and 2060, respectively. Electric or hydrogen-fueled aviation is not considered, as a conservative assumption based on expected long lead times delaying a significant market penetration.

Although shipping demand is projected to increase by 50\% to 2050, efficiency measures may counteract this to result in 0-30\% end energy demand increase, depending on the scenario \cite{IRENA2021}. We assume that efficiency measures are stronger towards 2060, resulting in a 20\% fuel demand increase compared to the base level, for both 2040 and 2060. IRENA \cite{IRENA2021} assumes an about 25\% share of hydrogen-based fuels (ammonia and hydrogen) for 2040, and about 55\% in 2050 in an ambitious scenario, with the rest being liquid or gaseous hydrocarbons (except very minor electrification). We thereby assume 25\% and 70\% hydrogen in 2040 and 2060, respectively, with the rest being liquid hydrocarbons. Electrification of shipping is not considered.

The total fuel demand in EU road transport consists of 64\% passenger road transport and 36\% freight road transport \cite{Mantzos2017}. Passenger and freight road transport services (i.e.\ passenger-km and ton-km) are projected to increase by 15\% and 24\% in 2040 as well as 20\% and 33\% in 2050 compared to 2018 \cite{JRC2019}. We assume increases of 25\% and 40\% for 2060. The freight share increases to 44\% and is assumed to have 40\% EVs and 60\% FCEVs by 2060. This results in 34\% and 68\% EVs in 2040 and 2060, respectively, and 5\% and 32\% FCEVs in 2040 and 2060, respectively. These assumptions are in line with a ban on new ICEVs by 2035 \cite{EuropeanParliament2021,Morfeldt2021}.

\begin{figure}[ht]
    \centering
    \includegraphics[width=0.4\textwidth]{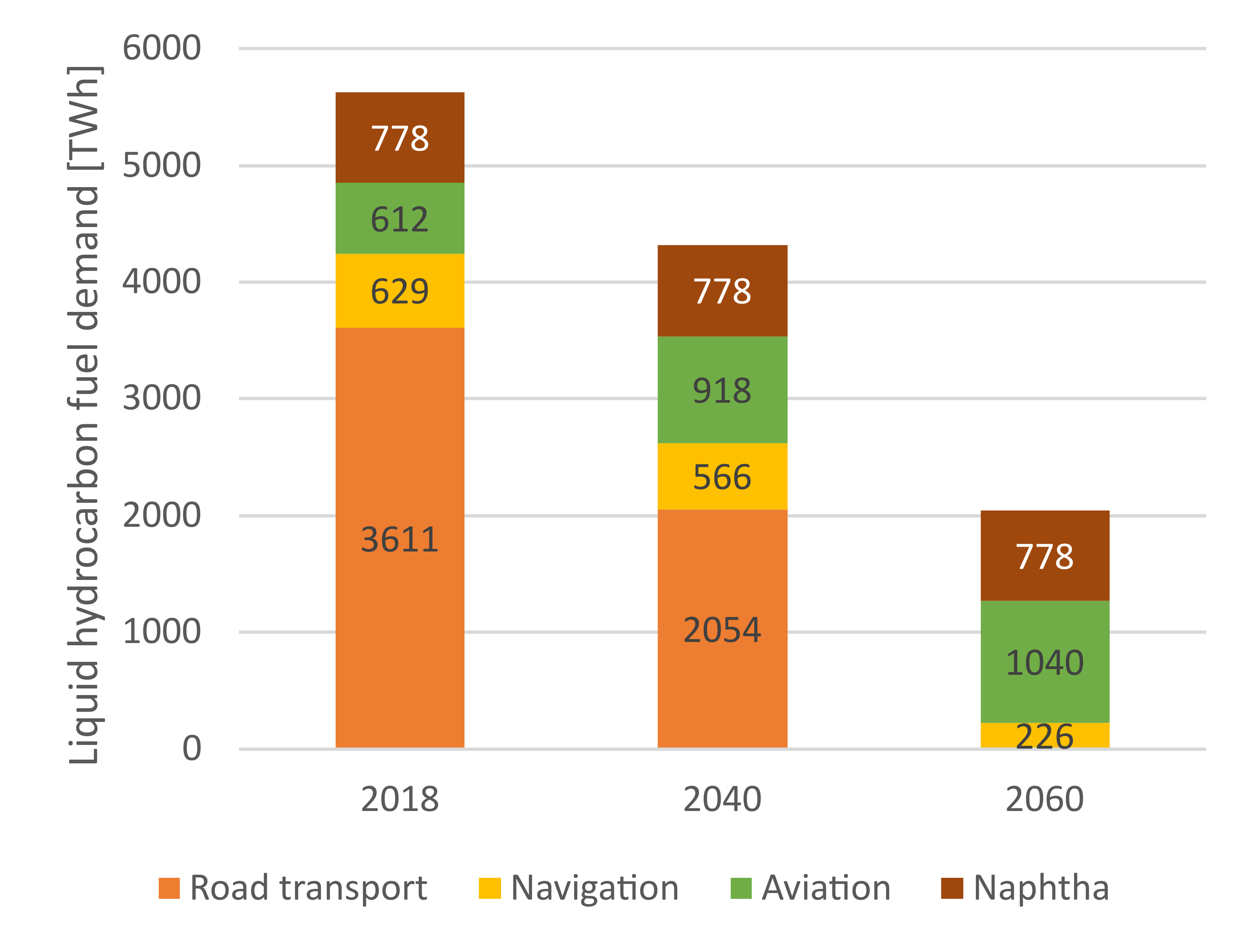}
    \caption{Assumed liquid hydrocarbon fuel demand in the years 2018 \cite{JRC2019}, 2040 and 2060.}
    \label{fig:fuelDemand}
\end{figure}

Total initial transport fuel demand (2018) amounts to 4851 TWh and includes road transport (incl. non-electric trains), domestic and international aviation and navigation (Figure~\ref{fig:fuelDemand}). The liquid fuel demand for industry feedstock (naphtha) is added on top of this and amounts to 778 TWh (held constant across years). The resulting total liquid fuel demand for transport amounts to 3444 TWh in 2040 (4223 TWh incl. naphtha, corresponding to \textasciitilde30\% of total primary energy demand) and 1233 TWh in 2060 (2011 TWh incl. naphtha, or \textasciitilde15\% of total primary energy demand). The inclusion of naphtha is justified as it is a part of the product mix from the Fischer-Tropsch process.

Steel production is assumed to be increasingly performed with hydrogen as a reduction agent (Direct Reduced Iron, DRI) and Electric Arc Furnaces (EAF), with 40\% in 2040 and 100\% in 2060. The share of scrap steel increases from 40\% today to 60\% 2040 and 70\% 2060.

The space heating demand is assumed to decrease by 16\% by 2040 and 29\% by 2060, through efficiency improvements in buildings. The district heating share is assumed at current levels for each country \cite{Zeyen2021}.

Industrial heat is divided into three segments: low, medium and high temperature. In the low and medium temperature segments, biomass is an option, whereas methane is an option in all three. Direct electrification is an option in the low temperature (process steam) segment, while heat pumps are excluded in the base case. Thus, solid biomass competes for producing industrial process steam with electric boilers and methane boilers, and for producing medium temperature process heat with methane.

\subsection{Scenarios}
The scenarios are varied in four dimensions: target year, biomass availability and carbon sequestration capacity, which have been identified as having a large influence on outcomes \cite{Blanco2018}, as well as liquid biofuel quota.

Two target years are analysed, namely 2040 and 2060. These target years are connected to different CO$_2$ emission targets, with an 80\% reduction in 2040 compared to 1990 and a 105\% reduction in 2060 (i.e.\ a net-negative target to represent the long-term need to remove carbon dioxide from the atmosphere; Sweden already has a net-negative target for after 2045 \cite{MinistryoftheEnvironmentandEnergy2018}.) Other than that, technology costs and efficiencies differ between the years, as presented in the supplementary information. For both years, it is assumed that conventional and renewable capacities existing in 2020 still exist in 2040 and 2060, unless they have reached the end of their life time. The two years are not interlinked, i.e.\ capacities built in scenarios for 2040 are not considered in scenarios for 2060.

\begin{table}[ht]
    \centering
    \begin{footnotesize}
    \begin{tabular}{l c c}
    \toprule
                      & Biomass & CS\\
                      & TWh & MtCO$_2$/a\\
                      \midrule 
High bio, low CS & 3608 & 0 | 400 \\
High bio, high CS & 3608  & 1500 \\
Low bio, low CS & 874 & 0 | 400\\
Low bio, high CS & 874 & 1500\\
\bottomrule
    \end{tabular}
    \caption{Biomass and carbon sequestration (CS) potentials assumed in the scenarios, for 2040 and 2060, resulting in eight base scenarios. In 2040, no carbon sequestration is assumed in the low CS scenarios, while in 2060 400 MtCO$_2$ is assumed because it is close to the minimum necessary in order to be able to achieve a net-negative target. For each of the base scenarios, different biofuel mandates of equal to 0\%, above 20\%, 50\% and 100\%, as well as no mandate (i.e.\ free optimisation) are assessed.}
    \label{tab:scenarios}
    \end{footnotesize}
\end{table}

The biomass availability is varied as presented in Section \ref{sec:biomass}. The carbon sequestration capacity is assumed at either 0 or 1500 MtCO$_2$/year for 2040, and 400 or 1500 MtCO$_2$/year for 2060, as summarised in Table \ref{tab:scenarios}. The lower end represents (close to) the least amount necessary to reach the set target, while the upper end is high enough to never be reached in the scenarios, i.e.\ it does not set an active constraint.

Production of liquid biofuels $\pi_{fu}$ is forced as a function of the share $\alpha \in [0,1]$ of the total set demand $\delta_{fu,s}$ for each sector $s$: (i) liquid fuels in the transport $trp$ (including land-based transport, marine and aviation) and (ii) industry $ind$.

\begin{equation}
    \pi_{fu} \ge \alpha \sum \delta_{fu,s} \ \  \forall s \in \{ind, trp \} 
\end{equation}

The total energy system cost as well as the estimated liquid fuel cost (\ref{sec:estimateFuelCost}) serve as metrics for the analysis.

\subsection{Sensitivity analysis}\label{sec:sensitivityMethod}
The sensitivity analysis is performed through runs of all the combinations of optimistic and pessimistic parameter value combinations outlined in Table \ref{tab:sensitivityranges}. All parameter values for a group are set to either pessimistic or optimistic, i.e.\ for example all Fischer-Tropsch-related parameters including for Biomass to Liquid and Electrofuels. This results in 2$^6$=64 combinations. These are run for each of the four main scenarios for 2060, with the model temporal resolution reduced to 237 representative time-steps per year (instead of 8760 with an hourly resolution) due to computational restrictions. A lower temporal resolution tends to slightly overestimate the biofuel share among liquid fuels.

\begin{table}[ht]\centering
\begin{footnotesize}
\begin{tabular}{llllccc}
\toprule
                  &                     &                     &           & Optimistic & Base & Pessimistic \\
                  \midrule 
Fischer-Tropsch   & Biomass to Liquid   & Investment cost     & €/kW      & 1500       & 2000 & 2500        \\
                  &                     & Efficiency          &           & 0.5        & 0.45 & 0.35        \\
                  & Electrofuels     & Investment cost     & €/kW      & 675        & 900  & 1125        \\
                  &                     & Efficiency          &           & 0.9        & 0.75 & 0.6         \\
                  \midrule 
Electrolyser      &                     & Investment cost     & €/kW      & 150        & 250  & 400         \\
                  &                     & Efficiency          &           & 0.8        & 0.75 & 0.7         \\
                  \midrule 
Carbon capture    & CHP                 & Cost                & €/ktCO$_2$/h & 1600       & 2000 & 2800        \\
                  & Industry            & Cost                & €/ktCO$_2$/h & 1400       & 1800 & 2400        \\
                  & DAC                 & Cost                & €/ktCO$_2$/h & 3000       & 4000 & 7000        \\
                  \midrule 
\multicolumn{2}{l}{Carbon sequestration}      & Cost                & €/tCO$_2$    & 10         & 20   & 50          \\
\midrule  
Fossils           & Oil                 & Price               & €/MWh     & 37.5       & 50   & 62.5        \\
                  & Gas                 & Price               & €/MWh     & 15         & 20   & 25          \\
                  \midrule  
Biomass           &                     & Import price (base) & €/MWh     & 36         & 54   & 72          \\
\bottomrule
\end{tabular}
    \caption{Assumed sensitivity ranges of key parameters directly relevant to liquid fuel supply in 2060. Ranges from DEA for 2050 \cite{DEA2021}, except for BtL, carbon sequestration, oil, gas and biomass imports, which are varied $\pm$25\%, except for the BtL efficiency (range based on literature) and carbon sequestration cost (assumed to have a higher cost variability). BtL includes the gasification unit as well as the FT-process.}
    \label{tab:sensitivityranges}
\end{footnotesize}
\end{table}

\section{Results}
In this section, we present the resulting fuel supply and solid biomass usage in the main scenarios without fuel mandates, and then we assess the effect of enforcing biofuel mandates on the system and show why such mandates risk increasing energy system costs substantially.

\subsection{Fuel and electricity supply in the base scenarios without biofuel mandates}
\begin{figure}[ht]
    \centering
    \begin{tikzpicture}
    \node (image) at (0,0) {
       \includegraphics[trim=60 50 60 70, clip, width=0.9\textwidth]{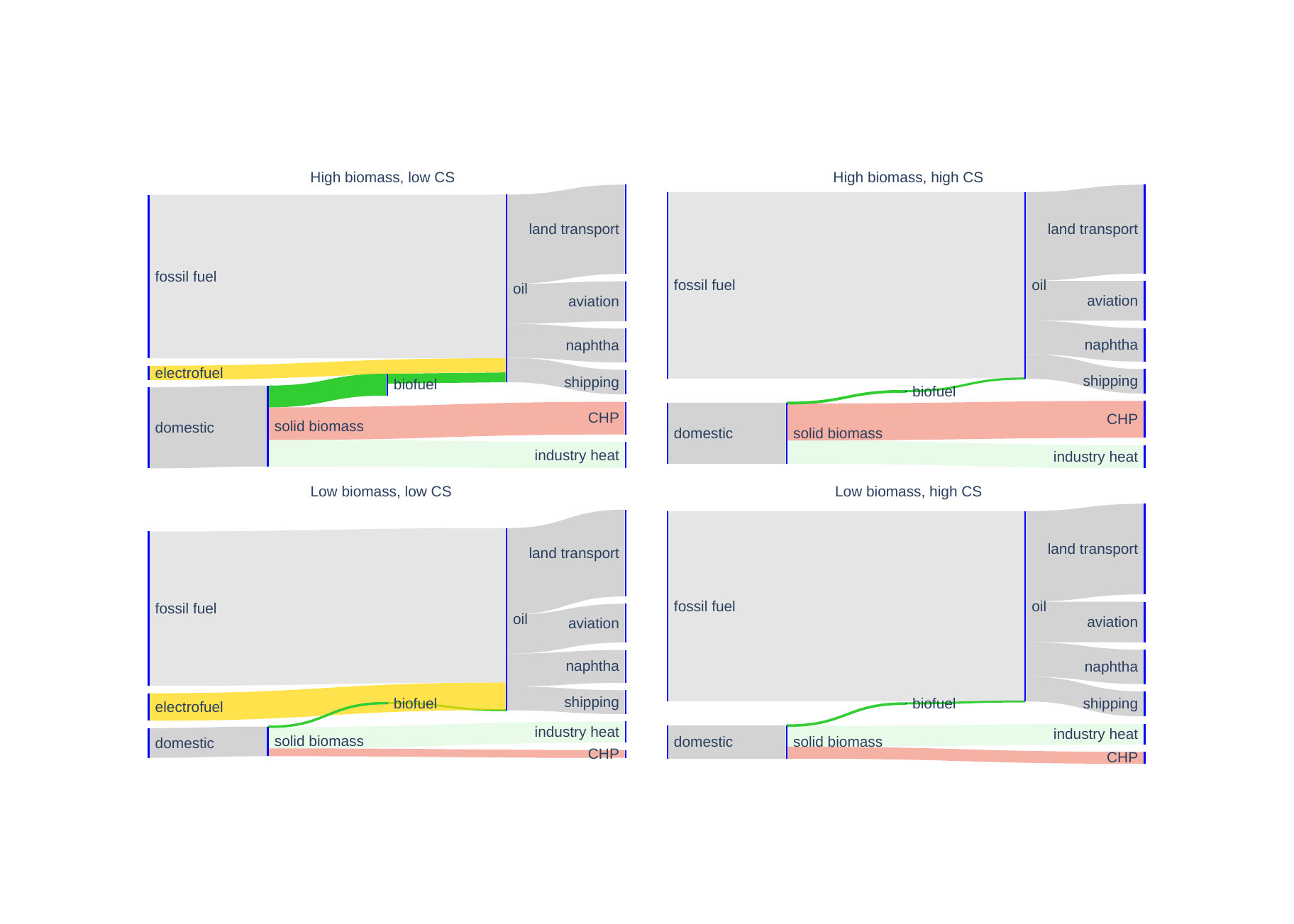}
    };
    \begin{scope}
        \node[left,black,rotate=90] at (-2.3,3.5){\scriptsize 4.32 PWh};
        \node[left,black,rotate=90] at (5.4,3.2){\scriptsize 4.32};
        \node[left,black,rotate=90] at (5.4,-1.5){\scriptsize 4.32};
        \node[left,black,rotate=90] at (-2.3,-1.6){\scriptsize 4.32};
        \node[left,black,rotate=90] at (-5.85,1.55){\scriptsize 1.86 PWh};
        \node[left,black,rotate=90] at (1.85,1.15){\scriptsize 1.41};
        \node[left,black,rotate=90] at (1.85,-3.45){\scriptsize 0.76};
        \node[left,black,rotate=90] at (-5.85,-3.43){\scriptsize 0.70};
    \end{scope}
    \end{tikzpicture}
    \caption{Sankey diagram of fuel supply and solid biomass usage in the base scenarios for 2040. Naphtha is used as a feedstock in industry.}
    \label{fig:sankey2040}
\end{figure}
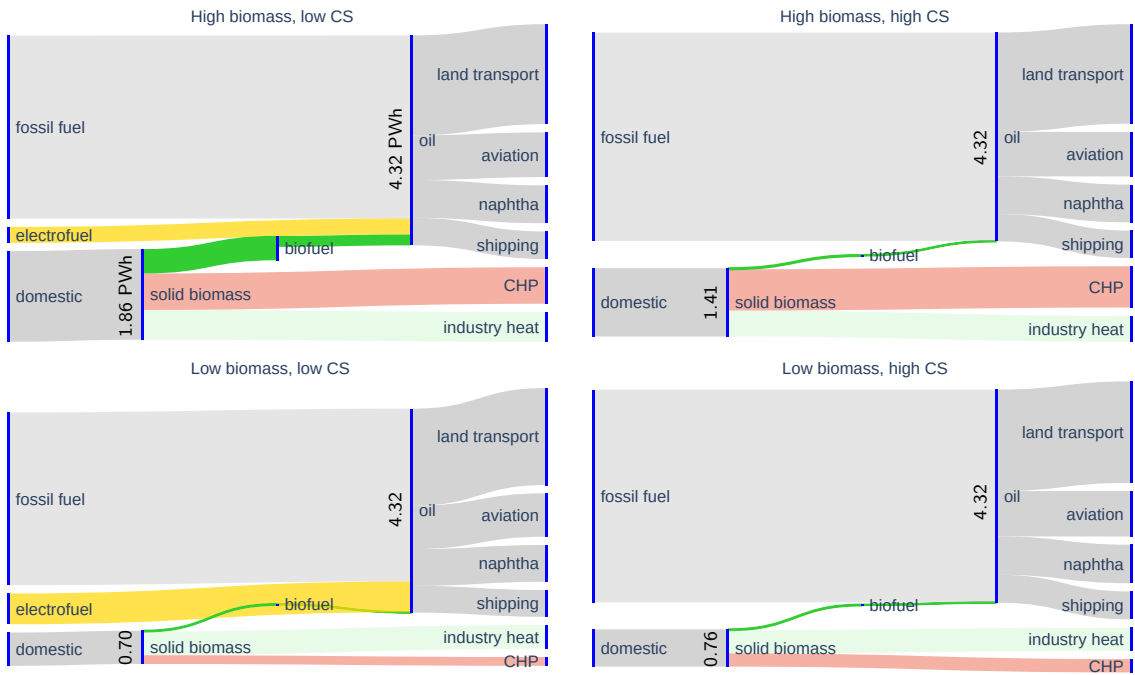

In all 2040 base scenarios (i.e.\ without a biofuel mandate), liquid fuel demand is dominated by fossil fuels (Figure~\ref{fig:sankey2040}). The reason is that there are more cost-effective abatement options to achieve an overall -80\% CO$_2$ emission reduction in the energy system. For example, the resulting electricity supply is almost fully supplied by non-fossil energy sources (mainly renewables, for which the resulting capacities are shown in Table \ref{tab:VREcapacities}). Also, electrification of transport, heat and industry contributes to decreasing emissions, and carbon capture is to some extent used in industry. With no carbon sequestration available, 13\% of the liquid fuel demand is covered by renewable fuels (5\% biofuels and 8\% electrofuels) in the optimal case if there is ample domestic biomass available, and 15\% electrofuels emerge if domestic biomass is scarce.

\begin{figure}[ht]
    \centering
    \includegraphics[width=1.0\textwidth]{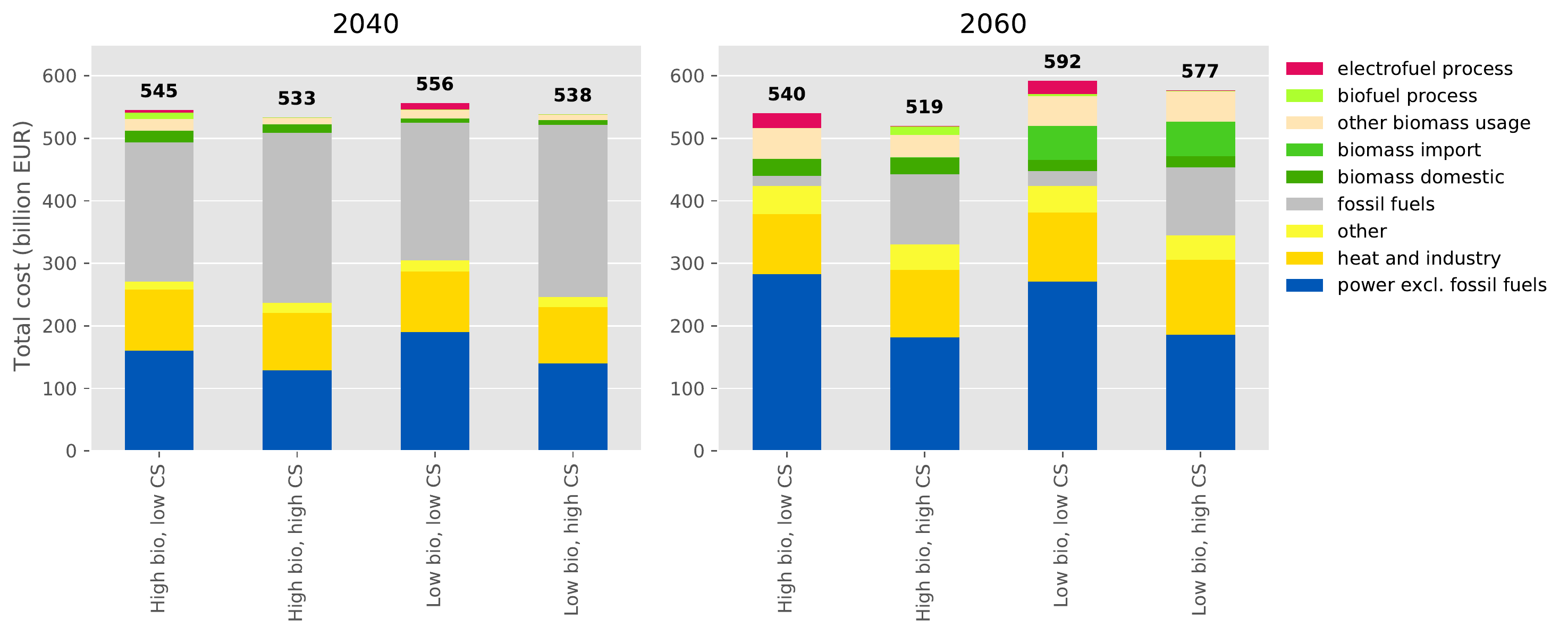}
    \caption{Total system cost [billion~\texteuro] in the 2040 and 2060 scenarios at different biomass and carbon sequestration availability, without biofuel mandates.}
    \label{fig:optimumResults}
\end{figure}

\begin{table}[]
    \centering
    \begin{footnotesize}
    \begin{tabular}{l c c c}
    \toprule \relax
        [GW$_p$] &  2020 & 2040 & 2060\\
        \midrule 
        Solar PV & 161 & 905 - 1269 & 1559 - 2776\\
        Onshore wind & 183 & 754 - 1073 & 914 - 1088\\
        Offshore wind & 25 & 166 - 322 & 290 - 696\\
        \bottomrule
    \end{tabular}
    \caption{Resulting VRE capacities in the base scenarios for 2040 and 2060, compared to values for Europe in 2020 \cite{IRENA2021}. The upper end of the resulting capacities is in scenarios with low carbon sequestration and low biomass.}
    \label{tab:VREcapacities}
    \end{footnotesize}
\end{table}

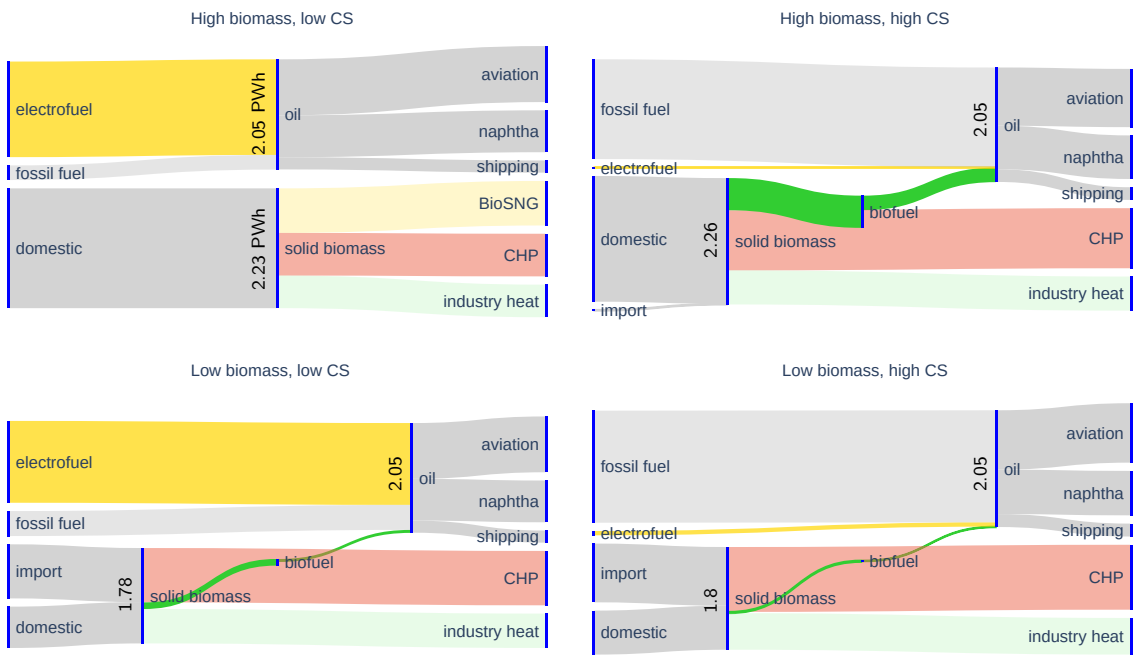
\begin{figure}[ht]
    \centering
    \begin{tikzpicture}
    \node (image) at (0,0) {
    \includegraphics[trim=60 50 60 70, clip, width=0.9\textwidth]{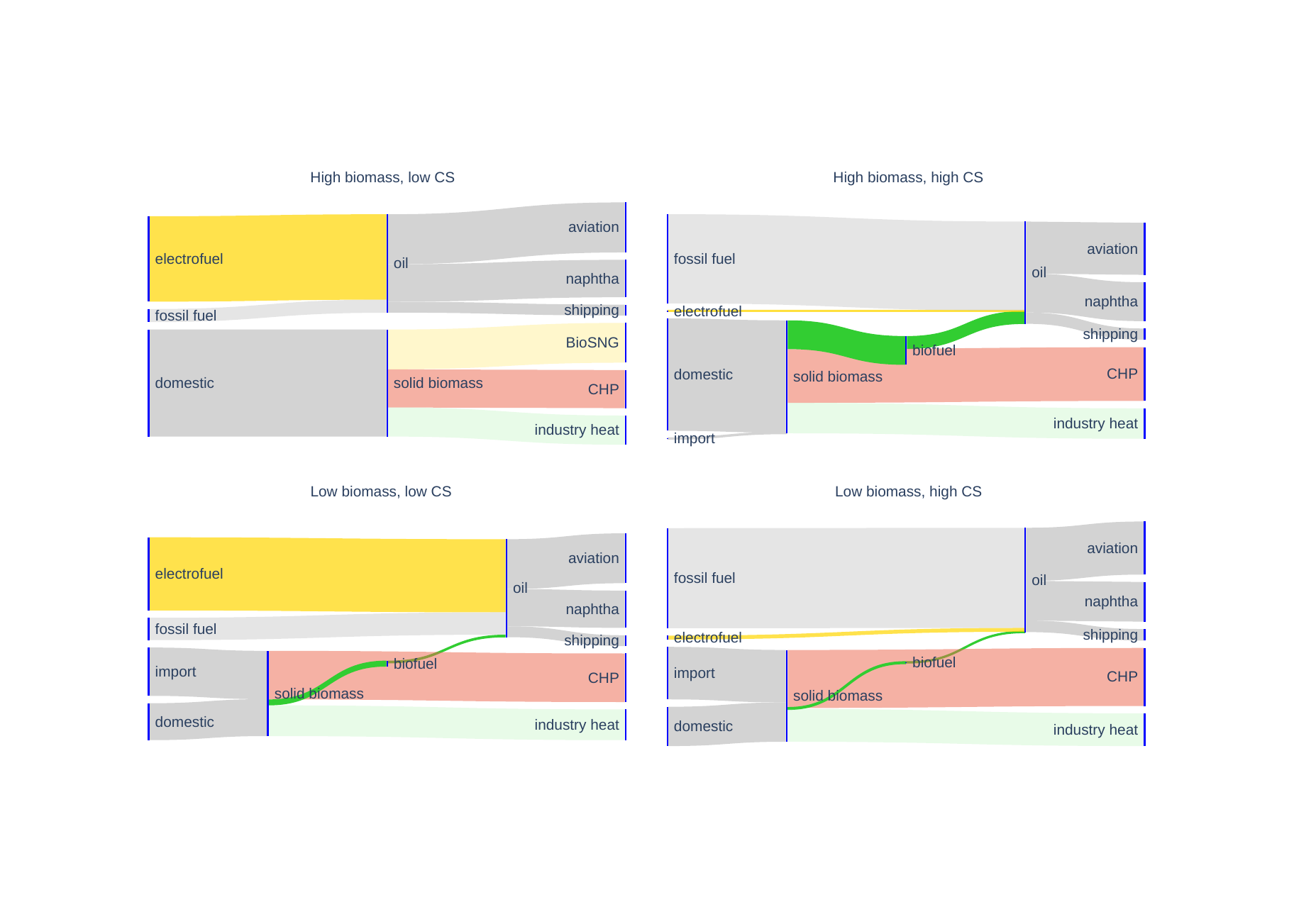}
    };
    \begin{scope}
        \node[left,black,rotate=90] at (-4.1,4){\scriptsize 2.05 PWh};
        \node[left,black,rotate=90] at (5.4,3.6){\scriptsize 2.05};
        \node[left,black,rotate=90] at (5.4,-1.1){\scriptsize 2.05};
        \node[left,black,rotate=90] at (-2.3,-1.1){\scriptsize 2.05};
        \node[left,black,rotate=90] at (-4.1,2.2){\scriptsize 2.23 PWh};
        \node[left,black,rotate=90] at (1.85,2){\scriptsize 2.26};
        \node[left,black,rotate=90] at (1.85,-2.85){\scriptsize 1.8};
        \node[left,black,rotate=90] at (-5.85,-2.7){\scriptsize 1.78};
    \end{scope}
    \end{tikzpicture}
    \caption{Sankey diagram of fuel supply and solid biomass usage in the base scenarios for 2060. Naphtha is used as a feedstock in industry.}
    \label{fig:sankey2060}
\end{figure}

In the 2060 base scenarios, the liquid fuel supply differs substantially between being dominated by electrofuels if carbon sequestration is scarce and by fossil fuels if there is ample carbon sequestration capacity available (Figure~\ref{fig:sankey2060}). With little carbon sequestration, the total electricity supply doubles compared to in the 2040 scenarios, while with ample carbon sequestration it increases by 30\%. In both cases, the additional electricity is covered mainly by solar PV and offshore wind power and is mainly used for applications which today rely on non-electric primary sources, i.e.\ supplying industry, heat and transport either directly with electricity or via producing hydrogen or methane which are used in those sectors.

Solid biomass is most cost-effectively used for CHP and industrial heat to varying degrees depending on the scenario, and with ample domestic biomass in 2060 also for producing some BioSNG. Biofuels make up a minor part of the biomass usage in all of the optimal cases (Figures \ref{fig:sankey2040} and \ref{fig:sankey2060}).

Now, optimal results may be rather sensitive to small perturbations in the system \cite{DeCarolis2016,Neumann2021} and therefore such results need to be handled with care. The question is: to which extent does a diversion from the optimal biomass usage and fuel supply affect system costs?

\clearpage

\subsection{How is the system affected when biofuel mandates are introduced?}
\begin{figure}[ht]
    \centering
    \includegraphics[width=1.0\textwidth]{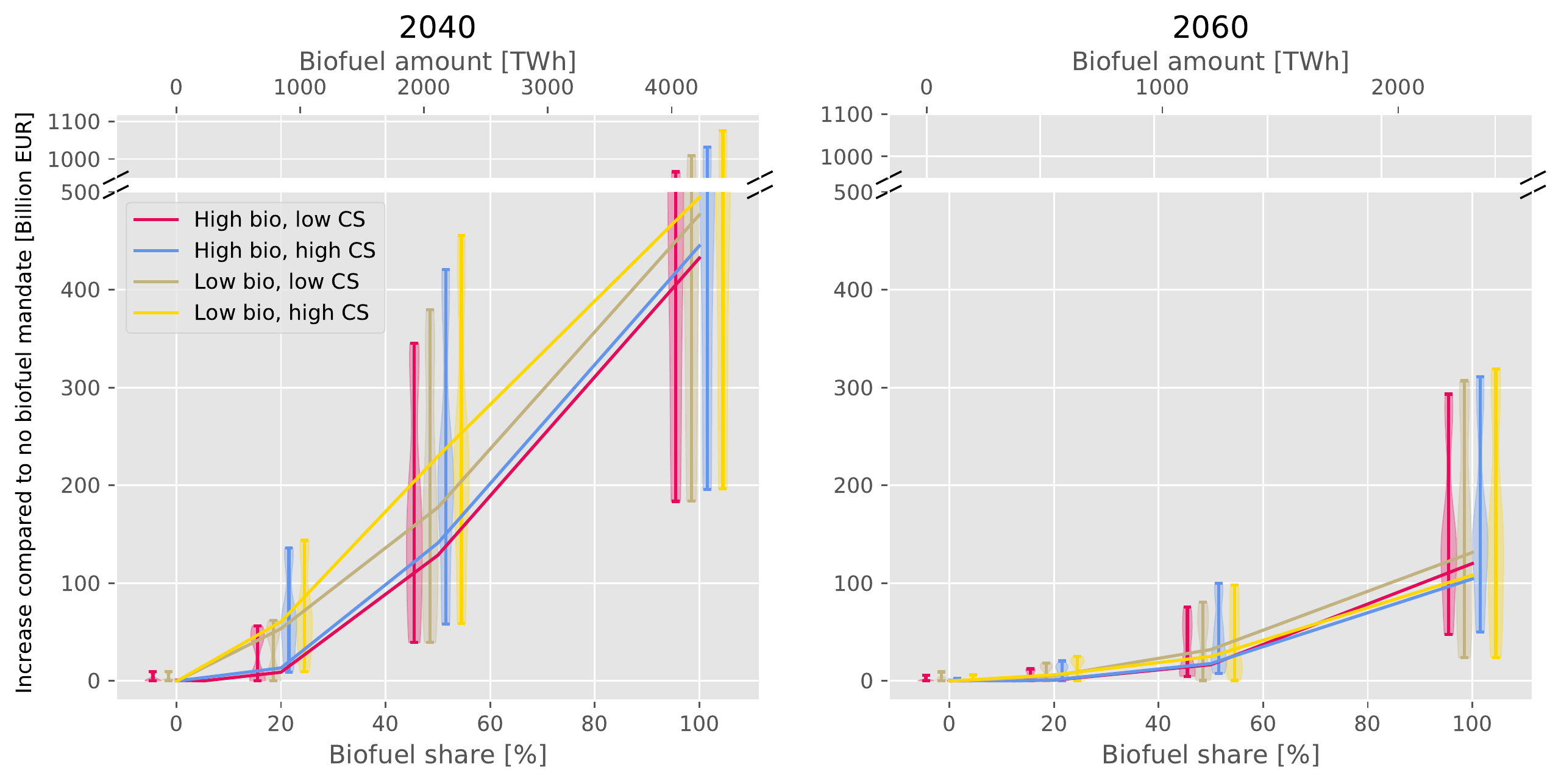}
    \caption{Total energy system cost increase compared to without a biofuel mandate. The vertical lines show the distribution from the sensitivity analysis for each biofuel share.}
    \label{fig:costincrease}
\end{figure}

The cost increase due to biofuel mandates in the different scenarios is shown in Figure~\ref{fig:costincrease}, both for the medium term (2040) and long term (2060). The figure also shows violin plots of the span and distribution of results from the parameter sensitivity analysis (Section \ref{sec:sensitivityResults}). 

A clear and general trend of increasing costs with increasing mandates can be observed, which is also robust to parameter variations. The cost increase due to biofuel mandates is higher in 2040 than in 2060, due to a higher liquid fuel demand and a lower emissions target in 2040. 

The remaining demand for liquid hydrocarbon fuels is among the most costly parts of the system to directly replace with renewables. Thus, at a -80\%-target in 2040, biofuel mandates substitute oil for much more costly biofuels. The more ambitious emissions target of -105\% in 2060 requires measures to be taken also for the liquid fuel supply, and thus the least-cost abatement option is more expensive (either electrofuels or fossil fuels combined with CCS), while the value of the competing usage of biomass for CHP and industry heat combined with carbon capture also increases.

Figure~\ref{fig:costincrease} also reveals that both the domestic biomass supply as well as the carbon sequestration capacity have a clear impact on the cost of biofuel mandates. The domestic biomass residues display low costs, and thus a lower supply of these leads to more costly imported biomass being needed when pursing higher biofuel mandates or more ambitious emission targets. The availability of carbon sequestration enables more fossil fuels to be used, which are substantially lower cost than biofuels. Therefore, biofuel mandates increase costs more in scenarios with a low biomass supply and with high carbon sequestration capacities. 

Figures~\ref{fig:costs2040} and~\ref{fig:costs2060} show the estimated costs broken down by fuel type, costs from fossil liquid fuel emissions caused elsewhere in the system, and cost increases in other parts of the system due to redirecting biomass usage to biofuel production. See \ref{sec:estimateFuelCost} for the method used to calculate these values. For the analysis in sections \ref{sec:2040results} and \ref{sec:2060results}, the fossil fuel emission cost is added to the fossil liquid fuel cost.

\subsubsection{Are biofuels a cost-effective transitional solution?} \label{sec:2040results}
Biofuels are sometimes put forward as a transitional solution to reduce the emissions of the transport sector until electrification achieves high shares \cite{Millinger2019,Riksdagen2018}. Despite the fact that a substantial electrification of transport was assumed, the fuel demand in 2040 is estimated at 71\% of that in 2020 and corresponds to \textasciitilde30\% of the resulting primary energy demand.

\begin{figure}[ht]
    \centering
    \includegraphics[width=1.0\textwidth]{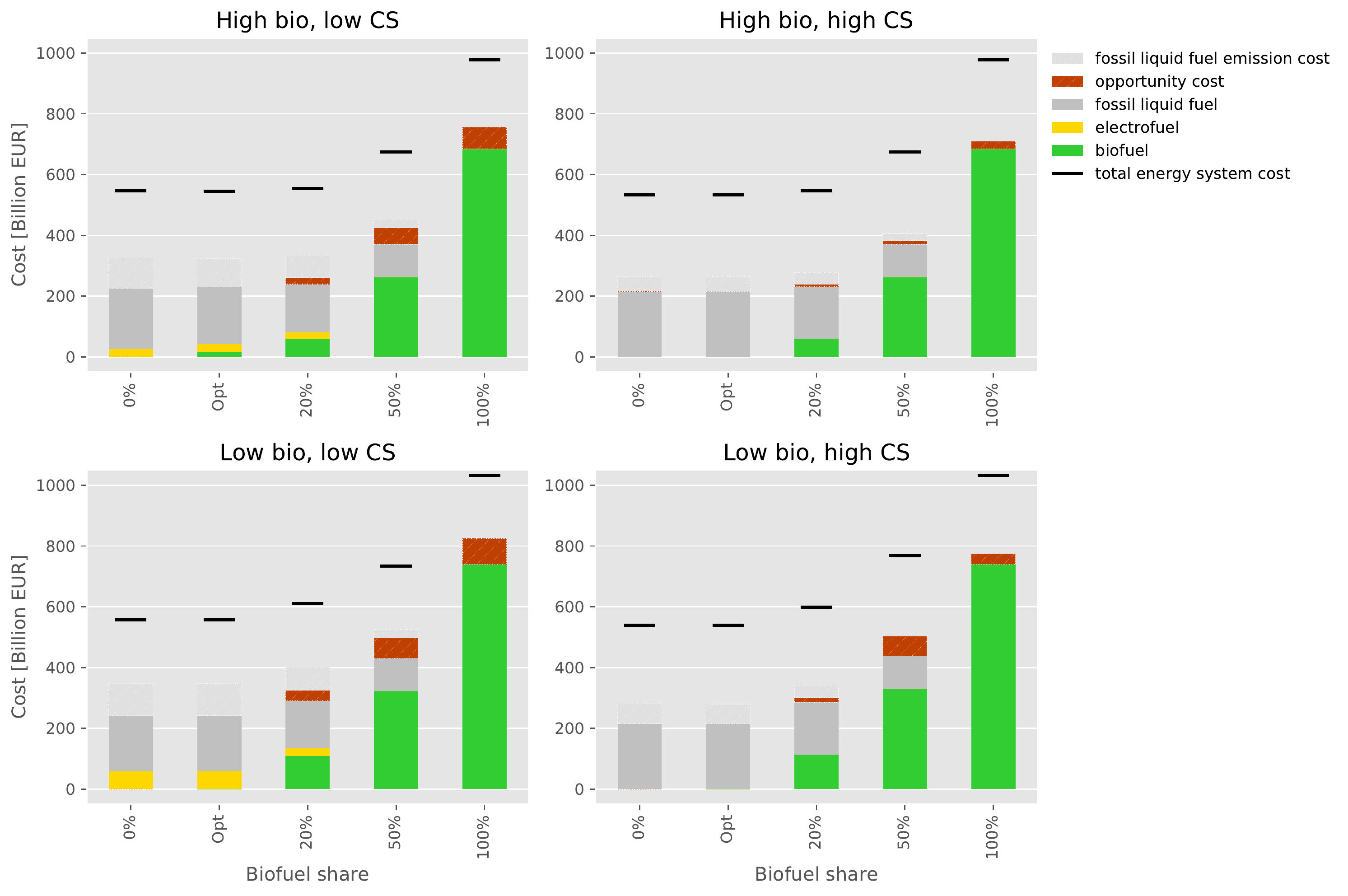}
    \caption{Liquid hydrocarbon fuel cost [billion~\texteuro /year] in the 2040 scenarios when pushing biofuels into the liquid fuel mix, at different biomass and carbon sequestration availability. The opportunity cost denotes the additional cost induced elsewhere in the energy system when redirecting biomass to biofuel production, and the estimated fossil liquid fuel emission cost likewise occurs in other parts of the energy system (both of which are striped.)}
    \label{fig:costs2040}
\end{figure}

The estimated cost of fuel supply (Figure~\ref{fig:costs2040}) in the optimal cases amounts to 265-347 billion~\texteuro if fossil fuel emission costs are included. A biofuel mandate of 20\% in 2040 results in a total cost increase of 8-60 billion~\texteuro \ (Figure~\ref{fig:costincrease}), where the lower value represents the case with high domestic supply of biomass. This corresponds to 2-21\% of the fuel cost without a mandate. Mandates of 50\% lead to cost increases of 128-229 billion~\texteuro, which corresponds to 39-82\% of the fuel cost without a mandate. A mandate of 100\% results in a very large cost increase of 432-494 billion~\texteuro. For comparison, the total cost of transport fuels in the EU in 2018 is estimated to 282 billion~\texteuro \ (\ref{sec:fuelCostToday}), so the cost increases are substantial.

\subsubsection{Are biofuels a cost-effective long-term solution?}\label{sec:2060results}
In the 2060 scenarios, the assumptions on electrification and switch to hydrogen in this work entail that the liquid fuel demand decreases to 36\% of that in 2018 and amounts to \textasciitilde15\% of the resulting primary energy demand. Thus, enforcing a biofuel mandate as a share of the demand has a smaller eﬀect on the total system cost than in 2040. 

\begin{figure}[ht]
    \centering
    \includegraphics[width=1.0\textwidth]{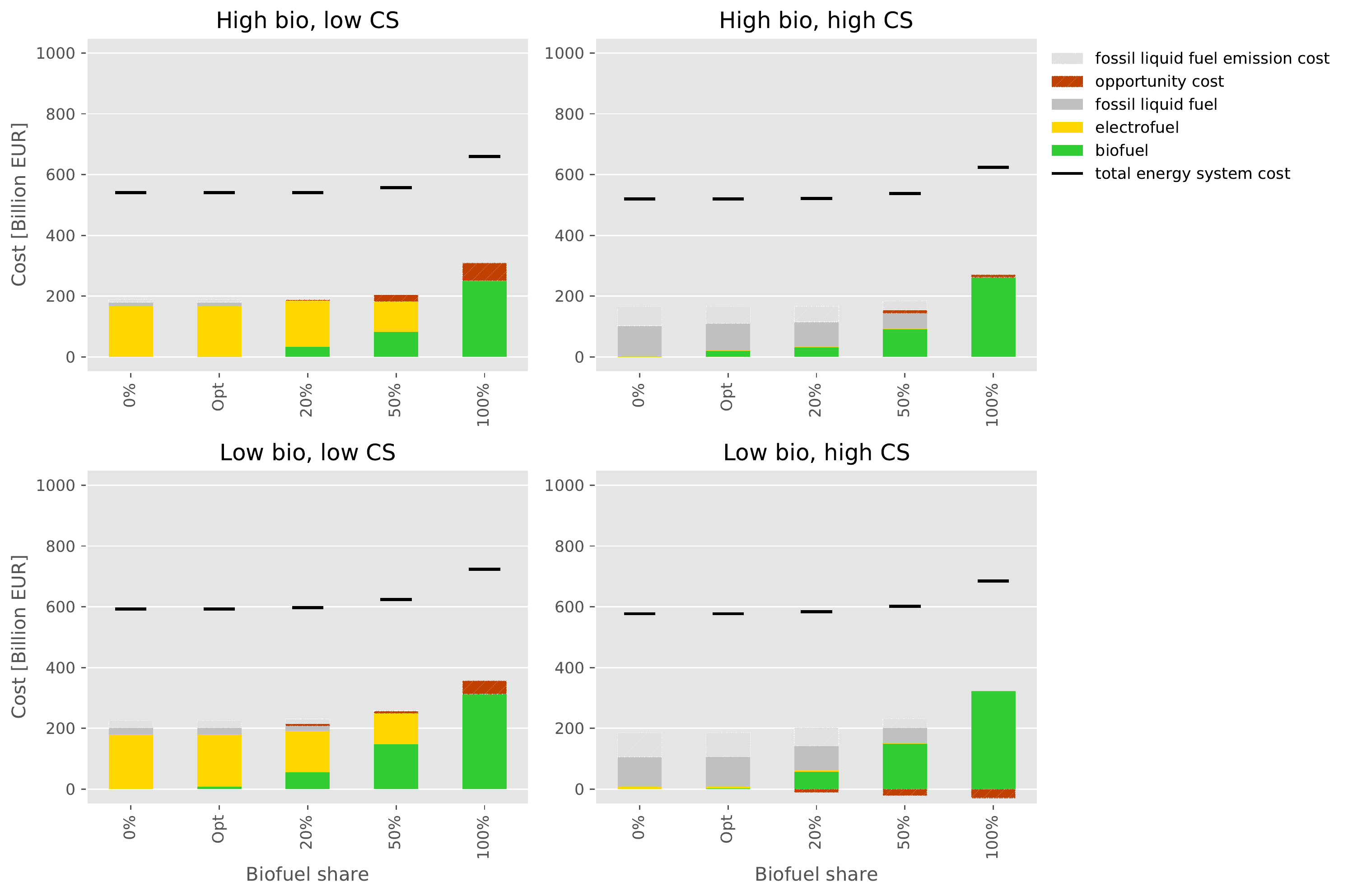}
    \caption{Liquid hydrocarbon fuel cost [billion~\texteuro /year] in the 2060 scenarios when pushing biofuels into the liquid fuel mix, at different biomass and carbon sequestration availability. The opportunity cost denotes the additional cost induced elsewhere in the energy system when redirecting biomass to biofuel production, and the estimated fossil liquid fuel emission cost likewise occurs in other part of the energy system (both of which are striped.)}
    \label{fig:costs2060}
\end{figure}

The cost of fuel supply in the optimal cases amounts to 166-225 billion~\texteuro \ if fossil fuel emission costs are included (Figure~\ref{fig:costs2060}). A 20\% biofuel mandate in 2060 results in a small cost increase of 0.8-6 billion~\texteuro \ (Figure~\ref{fig:costincrease}), which corresponds to 0.4-3\% of the fuel cost without a mandate. A 50\% mandate induces a cost increase of 16-32 billion~\texteuro, where the lower range is at an ample supply of domestic biomass. This corresponds to 8-14\% of the fuel cost without a mandate.

\subsubsection{What drives the cost increase when biofuel mandates are introduced?}
Redirecting biomass usage to biofuel production incurs several system effects. Generally, fuel costs increase due to biofuels being more expensive compared to both electrofuels and fossil fuels compensated with CCS (Figures \ref{fig:costs2040} and \ref{fig:costs2060}). The BtL process has a rather low conversion efficiency and a high investment cost, even though rather optimistic base values were chosen. Also, as there is a cost-supply curve for biomass, the more biomass is demanded, the higher the cost is, especially when expensive imports are needed.

Also, as the available biomass is used for fuel production it cannot be used for industrial heat and CHP, which instead are covered by other, more expensive options (direct electrification and methane). Thus, there is an opportunity cost of using solid biomass for liquid fuel production rather than for industrial heat and CHP (and at ample domestic biomass also for BioSNG), as other options there are more costly.

Furthermore, the potential for BECCUS is reduced, as a higher share of the biomass carbon can be captured in stationary combustion processes (assumed at 95\%) compared to when producing fuels, where only the carbon not ending up in the fuel can be captured (\textasciitilde 66\% at a conversion efficiency $\eta$=45\%). This has two effects: there is less biogenic carbon available for producing other hydrocarbons, and other, comparatively more expensive measures are needed to reduce emissions. Thus, more of e.g.\ biogas and power-to-methane are needed to decrease emissions in the system, at a higher cost.

The actual fuel cost in fact even decreases somewhat in some cases when biofuels from domestic biomass replace a share of the most expensive electrofuels, but there is always a total cost increase due to opportunity costs elsewhere in the system (Figure~\ref{fig:costs2060}). This reflects the cost of alternatives for supplying industry heat and CHP as well as carbon capture, and highlights the importance of covering the whole energy system when assessing biomass usage. Conversely, high biofuel mandates can also lead to slightly decreased costs for the rest of the system due to less CCS being needed to compensate for fossil fuels, but the overall system cost increases due to a substantially higher cost of fuel supply.

\subsection{Why are electrofuels preferred to biofuels when carbon sequestration capacity is scarce?}
The carbon used for producing the electrofuels stems from bioenergy with carbon capture, i.e.\ the carbon atoms are used twice in the system before being emitted to the atmosphere. This becomes important in the cases with little available carbon sequestration: fossil fuels cannot be compensated by CCS and DAC is more expensive, so renewable carbon atoms need to be utilized efficiently.

Also, solar and wind power are substantially more scalable resources than is biomass, and these serve as the main resources for producing hydrogen for electrofuels. Electrolysers can be utilised flexibly and thus run when electricity is cheaper. Thereby, they can also help solve integration issues at high variable renewable shares \cite{Ruhnau2021a,Ruggles2021,Bogdanov2021a}, and thus more variable renewables can be utilised cost-effectively.


\subsection{Sensitivity analysis}\label{sec:sensitivityResults}
The sensitivity of results to the 64 different parameter combinations of pessimistic and optimistic technology assumptions as outlined in Section \ref{sec:sensitivityMethod} is assessed.

The sensitivity on the total cost increase due to a biofuel mandate resulting from the varied parameters is substantial in all scenarios, but they consistently show an increasing trend with higher biofuel mandates leading to higher costs (Figure~\ref{fig:costincrease}). In 2040 the uncertainty is especially substantial, with up to a doubling of the total energy system cost at a 50\% biofuel mandate, and up to a trebling at a 100\% mandate.

\begin{figure}[ht]
    \centering
    \includegraphics[width=1.0\textwidth]{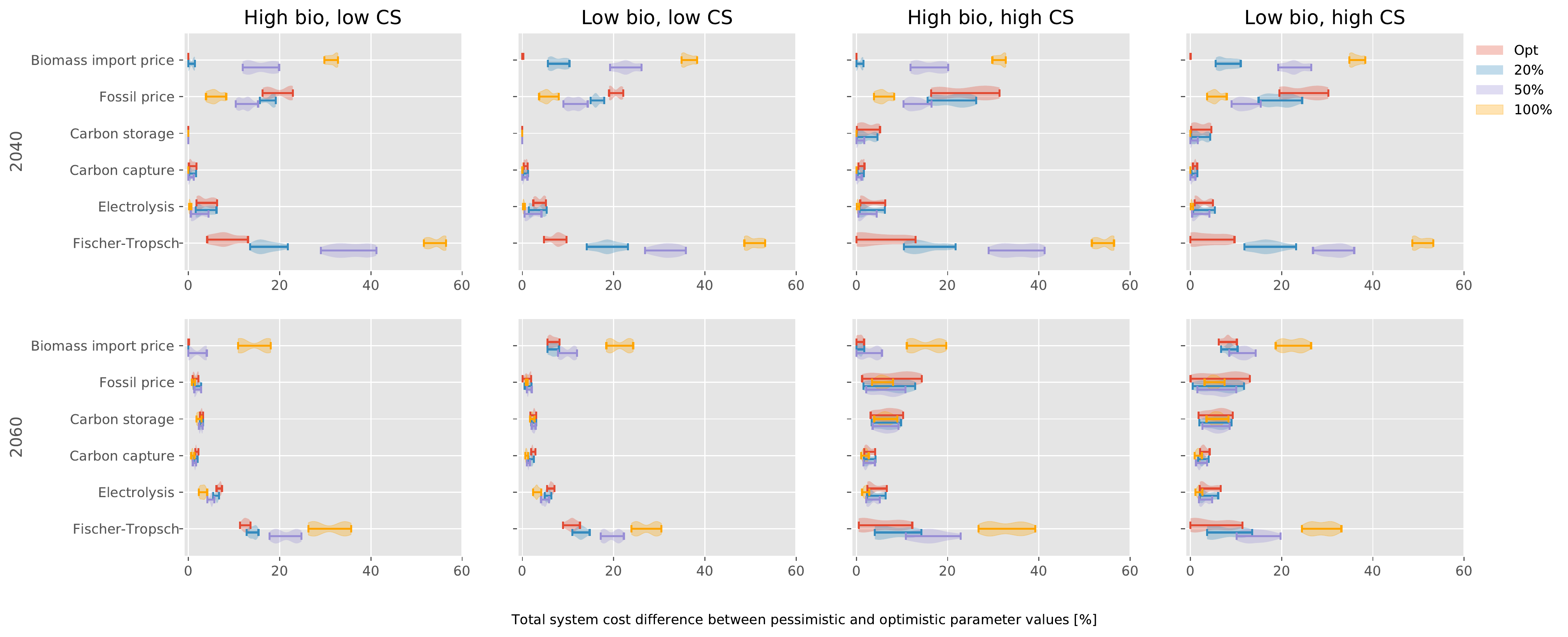}
    \caption{Violin plot of the parameter sensitivity of the system cost difference between pessimistic and optimistic parameter values for each extreme parameter value combination, at different biofuel shares in 2040 and 2060. This is shown in cases of low and high domestic biomass and low and high carbon storage potentials. The shaded areas show the distribution density of the outcome for a specific parameter when all other parameters are varied.}
    \label{fig:sensitivityViolin}
\end{figure}

The effect of the uncertainty of different parameters on the total energy system cost is shown in Figure~\ref{fig:sensitivityViolin}. The ranges show the uncertainty between pessimistic and optimistic parameter variations for each combination of the other parameters. For example, assuming high prices for fossil fuels in the optimal scenarios for 2040 compared to assuming low prices yield a minimum of 17\% total system cost difference, but it can yield a 33\% total system cost difference, depending on the values of the other parameters. Thus, a violin that is placed further to the right of the diagram means that the parameter in question is more important for the cost.

The parameter with the largest effect on the system cost is the investment cost and efficiency of Fischer-Tropsch, which affects both BtL and electrofuels. The effect increases with increasing mandates. The difference between pessimistic and optimistic values amounts to up to around 60\% of the total energy system cost at a 100\% biofuel mandate in 2040, or up to around 40\% in 2060.

The biomass import price has an effect especially at higher biofuel mandates, and more so in 2040 than in 2060, due to a higher fuel demand. In the optimal cases, the import price has barely any effect. This highlights that biomass imports do not play a role in any of those cases, despite the lower biomass import price being set to near current solid biomass prices. At increasing biofuel mandates however, biomass imports are needed.

The effect of the fossil fuel price uncertainty on the total energy system cost amounts to up to around 30\% in the optimal cases in 2040, and decreases with higher biofuel mandates. Fossil fuel prices and carbon storage costs have an effect in 2060 only if there is ample carbon sequestration capacity, since very limited fossil usage and carbon sequestration is allowed when carbon sequestration capacity is scarce.

In 2040, both carbon capture and carbon storage play a minor role and thus show little effect on total costs. The cost of carbon capture shows a minor effect also in 2060, while carbon storage cost uncertainty amounts to up to 13\% of the total cost if carbon sequestration capacity is large. The cost and efficiency of electrolysis shows an effect of up to 10\% on the total cost in both 2040 and 2060. The uncertainties of carbon storage, capture and electrolysis exhibit similar patterns across all cases and are thus independent of biofuel mandates as well as of the target year.

\section{Discussion}
In this section, we first compare the results to other studies, then discuss different factors affecting biofuel mandates.

\subsection{Comparison to other studies}
Most previous studies on the subject of optimal use of biomass in the energy system have been performed using models lacking the spatio-temporal resolution needed to represent VRE and electrolysis \cite{Iwanaga2021,Collins2017}, and therefore their interplay cannot be well represented in such studies.

Using the TIMES model with a low temporal resolution, Blanco et al.\ \cite{Blanco2018} assessed hydrogen usage options including electrofuels to find that carbon sequestration capacity (and thus the possibility to continue using fossil fuels), biomass availability (which limits the green CO$_2$ available for electrofuel production) as well as technology costs determine the competitiveness and potential of electrofuels. These findings align with the results presented here. Solid biomass is primarily used for hydrogen production (which was allowed but did not show up in our results) and electricity generation unless carbon sequestration is not allowed, in which case biofuel production ends up as the main solid biomass usage option, in contrast to our results.

Similarly to our results, Lehtveer et al.\ \cite{Lehtveer2019} find that a scarcity of carbon sequestration capacity is decisive for the competitiveness of electrofuels. They, however, find electrofuels to be less cost-competitive than biofuels in most cases. The contrast to the results in this work can at least partly be explained by substantially higher assumed costs for VRE and electrolysers (with many cost assumptions for VRE for 2050 being above current actual costs), while biomass and biofuel conversion costs are very similar.

Millinger et al.\ \cite{Millinger2021} assessed renewable fuel competition in Germany and found that electrofuels were cost-competitive to biofuels in the long term (2050), and necessary for achieving high renewable shares in transport, in line with the results presented here. The availability of excess renewable electricity and dwindling carbon sources were found to limit the electrofuel production potential, similar to in this study. 

Other energy system studies which include electrofuels were not found \cite{Leblanc2022,Bauer2020,Ahlgren2017,Azar2003}, which makes comparisons to this study difficult. Many IAM studies arrive at the result that high carbon sequestration capacities enable a continued high use of fossil fuels in transport, in which case biofuel usage is low \cite{Ahlgren2017,Bauer2020,Lehtveer2019}. The usage of biomass for fuel production rather than for bioelectricity is sometimes determined by whether it can be combined with CCS or not \cite{Leblanc2022,Bauer2020}. IAM studies sometimes arrive at a high penetration of biofuels and sometimes low \cite{Ahlgren2017,Bauer2020,Azar2003}, depending on e.g.\ if CCS is possible to combine with bioliquid production, but also other model-specific assumptions \cite{Bauer2020}. However, as electrofuels were not included, the results cannot be directly compared with this study. In contrast, Bogdanov et al.\ \cite{Bogdanov2021} performed an energy system study with a high spatio-temporal resolution which included electrofuels, but biofuels as well as BECCS were excluded and thus results cannot be compared to this study.

\subsection{Other factors affecting biofuel mandates}
Biomass residues are a limited resource, and the domestic biomass residues assumed to be available in this study could only cover a part of the liquid fuel demand, even if it were all used for biofuels and despite assuming an ambitious electrification of transport. Within these bounds however, some factors are discussed below, which may affect the competitiveness of using the limited biomass for producing biofuels, or are relevant to biofuel mandates.

\subsubsection{Biofuel production and resource base}\label{sec:discussionBiomass}
In the sensitivity analysis, the most sensitive parameter was found to be the cost and efficiency of Fischer-Tropsch processes. In the base case, these parameters were set to figures in the optimistic part of the range based on literature \cite{Millinger2017,Dimitriou2018,Larson2012}. Even if a further 25\% cost reduction is assumed, a 50\% biofuel mandate still always resulted in a cost increase compared to without a mandate.

Additional low-cost biomass resources, could increase the potential for cost-competitive biofuels. Estimates of the possible contribution from dedicated biomass plantations and increased usage of forest biomass span a wide range (see e.g.\ Mola-Yudego et al. \cite{Mola-Yudego2017}). We did not consider biomass from dedicated plantations explicitly but note that they may increase the total biomass potential substantially and may complement residue resources in locations where these cannot meet demand \cite{Cintas2021}. Short-rotation-coppice (SRC) has been put forward as a low-risk option to increase the biomass potential and at the same time contribute to other ecosystem services, but it is uncertain to which extent and to which cost this can be realised \cite{Englund2020}.

The main biomass residue sources that have not been included here are oil-rich food wastes and crops. The potential of oil-based rest-products is relatively low compared to the demand for liquid fuels \cite{Meisel2020}. The domestic potential for used cooking oil (UCO) in the EU has been estimated at 60 PJ \cite{VanGrinsven2020}, which would correspond to 0.3\% of the present EU transport fuel demand \cite{Mantzos2017}. Besides sustainability concerns \cite{Creutzig2015,Creutzig2016}, food crops used for conventional biofuels have low yields and are thus also more sensitive to price increases due to land scarcity \cite{Millinger2018a}. The reliance on food crops for bioenergy is being decreased in the EU \cite{EuropeanParliament2018}, but the allowance of dedicated crops is subject to continued case-specific negotiations within the new EU Common Agricultural Policy \cite{EUCAP2021}.

In this study, biomass supply \textit{costs} are used for the domestic biomass. However, especially for goods such as solid biomass that are tradable on the global market, prices may be substantially higher than these costs. This means that the actual fuel cost here is underestimated, and using market prices for biomass would thus increase the cost of biofuels.

The assumption that bioenergy systems included in the model are carbon neutral is a simplification judged to be appropriate considering the scope of the study and the assumption that only agriculture residues and organic waste are included as bioenergy feedstock. Residue extraction may influence the soil carbon stocks negatively, but some uses of residues and waste for energy can facilitate recirculation of carbon to soils and can also reduce methane emissions \cite{Oehmichen2021}.

\subsubsection{Limitations for electrofuel production and CCS}
Electrofuels rely on a low-carbon source of electricity for hydrogen electrolysis, and a renewable source of carbon, which are both limited today and may be so also in the foreseeable future \cite{Millinger2021,Ueckerdt2021}.

Achieving the high electricity generation and carbon capture and storage capacities in Europe required for negative emission scenarios is an unprecedented challenge \cite{Cherp2021}. If cheap low-carbon electricity (e.g.\ mainly wind and solar PV) turns out to diffuse slowly, it is difficult to achieve ambitious emission targets. If domestic capacities are limited, electrofuels may instead be imported from regions with high solar and/or wind potentials and less land constraints, possibly produced at a lower cost than domestically \cite{Hampp2021,Kan2022}.

The difference in total cost between the scenarios with low and high carbon sequestration capacity in this study is small, despite the large difference in fuel mix (Figure~\ref{fig:costs2060}) as well as in the resulting VRE capacities (Table~\ref{tab:VREcapacities}), and deserves further attention in future work. A continued reliance on fossil fuels which are compensated by CCS or other CDR measures involves risks and is subject to controversy \cite{Middleton2018,Strefler2021,Anderson2016,Bistline2021}.

In this study, only carbon capture of process emissions was included. Several other CDR options exist but were out of the scope of this study, notably land-based options such as afforestation and reforestation (A/R), biochar, soil carbon sequestration, and BECCS relying on other biomass sources than the ones included in this study. A/R can provide CDR via carbon sequestration in vegetation and soils while at the same time providing biomass for various purposes including bioenergy and BECCS. But A/R may also take forms that compete for land with bioenergy options, e.g. revegetation to establish natural ecosystems not intended for biomass harvest. A/R may also result in a higher GHG abatement than energy crops used to replace fossil fuels \cite{Chan2022}, but short-rotation forests may become cost-effective at higher carbon prices \cite{Hedenus2009}. An analysis of such competitions is out of the scope of this study.

\subsubsection{Electrification of transport and hydrocarbon demand}
The high costs and conversion losses involved when producing hydrocarbon fuels, regardless of pathway, render hydrocarbon fuels one of the most costly parts of the energy system. Substantial additional VRE capacities and biomass are needed to cover for this as indicated in the scenarios, despite high electrification shares of road transport. Hydrocarbon demand reduction is therefore essential.

An endogenous optimisation of the vehicle park in the transport sector was not included in this study, for several reasons: in the land transport sector, the choice of vehicle is not subject to a simple cost-minimisation, but is a function of e.g.\ agents' desire for flexibility for occasional longer journeys, status and availability of charging infrastructure, as well as being subject to combustion engine bans \cite{EuropeanParliament2021}, and vehicle fleet inertia \cite{Morfeldt2021}. The high cost and resource need of renewable liquid fuels found in this study, as well as decreasing costs and supreme efficiency of EVs compared to ICEVs (ICEVs run on electrofuels require \textasciitilde5-7 times the amount of electricity per kilometer \cite{Gustafsson2015}) render EVs a highly competitive option. Future studies with PyPSA-Eur-Sec are underway to assess this topic.

In navigation, substantial energy efficiency improvements were assumed. H$_2$ was assumed to cover 70\% of the energy demand by 2060 \cite{IRENA2021}. This is of course highly uncertain. In aviation, no H$_2$ or electrification was assumed. Higher shares of EVs or H$_2$ or an overall lower demand in these sectors would decrease the liquid hydrocarbon fuel demand, and thus decrease the cost of biofuel mandates. However, demand for aviation and navigation may also increase more than assumed here.

\subsubsection{Cost-competitiveness of biomass usage in industry and CHP}
The cost-competitiveness of biomass usage for process heat in industry would be affected by a cheaper than expected electrification of industrial process heat. Industrial heat pumps for steam generation could be a competitive option \cite{Jordan2019,Jordan2020}, but it is uncertain to which extent they are a viable option for process steam \cite{Madeddu2020}. In medium and high temperature applications, electrification is possible, but the uncertainty increases with the required temperature and options are currently in experimental or pilot stages of technological readiness \cite{Madeddu2020}. Many processes such as steel production were assumed to be electrified to a high extent in this study, but we were conservative with electric options for process heat due to the above uncertainties. A sensitivity run where heat pumps for process steam were included resulted in biomass still being preferred for process steam. This, however, depends on the assumed COP, which depends on temperature differences to the heat sink; this is process specific and outside of the scope to assess in more detail here.

Biomass or other flexibility options are needed for CHP especially during cold dark doldrums, when both heating and electricity are needed but solar PV and wind generate little, and heat-pumps are less efficient \cite[c.f.][]{Zeyen2021}. Biomass for CHP appears without carbon capture in the 2040 scenarios. In a sensitivity analysis where BECCS was turned off for all technologies, CHP still appeared in 2060, to a similar extent as with BECCS turned on. Thus, this flexibility option provides an important system service and is not only due to the higher potential for BECCS compared to when producing biofuels. Other flexibility options for generation, such as batteries as well as heat and hydrogen storage are included in this study. A back-up system relying on renewables and not fossils is necessary at more ambitious emission targets, and biomass turns out to be a potentially cost-effective candidate for this. Note that CHP requires a district heating grid, which was restricted to current shares on a country level.

\subsubsection{Policy and co-benefits?}
Pursuing a renewable fuel mandate which exceeds the future fuel demand in a shrinking market presents a risk of stranded assets for investors or the risk of a lock-in effect which increases the system cost.

Even though renewable fuel shares turned out low at an 80\% emission reduction, as other measures to achieve the targets were less costly, there are still arguments for supporting such options earlier. Since the development of renewable fuels requires several parts in the supply chain to function, it may take time to set up the necessary infrastructure and logistics.

This goes for biofuels, which require the mobilisation of currently unused biomass residues as well as the investment in costly biofuel production facilities. It goes just as well for electrofuels, which require large amounts of clean electricity and a carbon source, as well as costly production facilities \cite{Millinger2021}. Uncertainties along the value chain and regarding sustainability issues and future prices make investments risky. This hinders the development of renewable fuels, and thus directed policy may be warranted as a complement, even if a first-best cap-and-trade or tax policy is implemented for the whole energy system \cite{Bennear2007,Jaffe2005,Lehmann2013}. Such policies could include sector-specific targets, technology subsidies and fuel blending mandates.

A general view is that technologies should be supported to address two market failures: the external costs of GHG emissions as well as of learning effects, which lead to an underestimation of future benefits or that they are not appropriated by the investor \cite{Jaffe2005,Lehmann2013}. Does this apply to biofuels?

Although the conversion technology may need time to develop, the biofuel price depends on both the investment and the resource, in contrast to e.g.\ for VRE. Biomass scarcity and the competitiveness of other biomass usage options may lead to biomass price increases which surpass investment cost reductions achieved through learning effects. Thus, it can be questioned whether supporting advanced biofuels paves the way for a promising technology in terms of cost reduction potentials (see \cite{Azar2003} for a similar argument).

However, there may be co-benefits and spillover effects between BtL and electrofuels, since they are both based on the FT-process. Thus, also the electrofuel process may improve in terms of cost and efficiency if BtL improves. It is also possible to combine biofuel and electrofuel production (for instance as electrobiofuels where biofuels are produced with a hydrogen addition, thereby using the biomass carbon more efficiently \cite{Celebi2019,Millinger2021}), or to reuse biofuel facilities for electrofuel production. It may also stimulate a transition to producing renewable chemicals in biorefineries. Therefore, supporting biofuels to some extent may still be a sensible investment in terms of research and development, as well as for setting up value chains and stimulating the mobilisation of currently unused biomass residues.

Nevertheless, care needs to be taken to ensure that production is indeed able to switch over time as outlined above as well as to accommodate for a changing fuel mix, in order to avoid infrastructural lock-in effects. Institutional lock-ins related to actors with vested interests \cite{Seto2016} may also present a challenge in this regard, if biomass streams are first stimulated and then directed away from renewable fuel production \cite{Reid2020}.

\section{Conclusions}
This work focuses on the competition for liquid fuel supply under CO$_2$ emission reduction targets, and the effects of biofuel mandates on the cost of the future European energy system.

Results indicate that in the medium term (\textasciitilde2040) VRE and a partial electrification of transport, heat and industry would suffice to achieve a -80\% emission reduction target cost-effectively. This would allow for a continued use of fossil fuels for the remaining liquid fuel demand, which is among the most costly parts of the system to directly cover with renewables. Introducing a biofuel mandate of 20\% resulted in a cost increase corresponding to 2-21\% of the fuel cost without a mandate. A 50\% mandate resulted in a cost increase of 128-229 billion~\texteuro \ depending on the scenario, or 39-82\% of the fuel cost without a mandate, with an uncertainty range due to parameter variations of 40-470 billion~\texteuro.

In the long term (\textasciitilde2060), liquid fuel demand is expected to be substantially lower due to electrification, and at a negative emissions target (-105\%) liquid fuels must be either renewable or compensated by CDR. However, biomass use for industry and CHP allows for more carbon capture than when producing biofuels, and this enables carbon atoms to be used several times in the system. Also, CHP emerged as an important flexibility option for heat and electricity supply. Electrofuels based on captured biogenic carbon emerged as the main fuel if carbon sequestration availability was low, while fossil fuels compensated by BECCS emerged if carbon sequestration availability was high. Notably, the difference in total cost was merely 3-4\% between these two systems, but VRE capacities differed substantially, with e.g.\ solar PV capacities ranging between 1.6 and 2.8 TW$_p$ and offshore wind between 0.3 and 0.7 TW$_p$. In this case, a 50\% biofuel mandate increased the total energy system cost by 16-32 billion~\texteuro \ depending on the scenario, corresponding to 8-14\% of the fuel cost without a mandate, with an uncertainty range due to parameter variations of 3-140 billion~\texteuro.

We conclude that even low biofuel mandates risk increasing total energy system costs substantially, and that this cost increase is higher if biofuel mandates are employed in the short- to medium term. Renewable fuel mandates may be warranted to establish necessary supply chains and flexible renewable fuel systems in terms of both feedstocks and products in a changing and uncertain environment appear sensible. However, a shrinking fuel market due to electrification may also present a risk of stranded assets for investors if high renewable fuel mandates are pursued early on. Biofuel mandates were found to increase system costs across a range of parameter variations and scenarios. The cost drivers are: (i) high biomass costs due to scarcity, (ii) opportunity costs for competing usages of biomass for industry heat and combined heat and power with carbon capture, and (iii) lower scalability and generally higher cost for biofuels compared to other abatement options (electrofuels and fossil fuels combined with CDR).

\section{Acknowledgements}
This publication is the result of a project carried out within the collaborative research program Renewable transportation fuels and systems (Förnybara drivmedel och system), Project no. 50460-1. The project has been financed by the Swedish Energy Agency and f3 – Swedish Knowledge Centre for Renewable Transportation Fuels.

The computations and data handling were enabled by resources provided by the Swedish National Infrastructure for Computing (SNIC) at C3SE, partially funded by the Swedish Research Council through grant agreement no. 2018-05973.

We thank the project reference group for fruitful discussions: Åsa Håkansson (Preem), Raziyeh Khodayari (Energiföretagen), Eric Zinn (Göteborg Energi), Sven Hermansson (Södra) and Svante Axelsson (Fossilfritt Sverige).

Viktor Rehnberg, Fredrik Rönnvall, Per Fahlberg and Thomas Svedberg at C3SE are acknowledged for assistance concerning technical and implementational aspects in making the code run on the C3SE resources.

We also thank Daniel Johansson, Johannes Morfeldt and Olivia Cintas Sánchez for helpful discussions.

\appendix




\section{Carbon balances of fuels}
Solid biomass carbon dioxide uptake from atmosphere, with \%C$_{sb}$=50\%, e$_{sb}$=18 GJ/t, m$_{CO_2}$/m$_C$=44/12 (Eq. \ref{eq:cuptake}): 
\begin{equation}\label{eq:cuptake}
    \varepsilon_{at}^{sb} = -\%C_{sb} \cdot \frac{3.6}{e_{sb}} \cdot \frac{m_{CO_2}}{m_{C}}
\end{equation}

Liquid fuel carbon dioxide emission [tCO$_2$/MWh] at full combustion for diesel and methane based on -CH$_2$- simplification and $e_{CH_2}$=44 GJ/t LHV for diesel, and $e_{CH_4}$=50 GJ/t LHV for methane (Eq. \ref{eq:combustionemission}):
\begin{equation}\label{eq:combustionemission}
    \varepsilon_{fu} = \frac{3.6}{e_{CH_x}} \cdot \frac{m_{CO_2}}{m_{CH_x}}
\end{equation}

The carbon share ending up in the fuel $C_{fu}$ : $C_{in}$ can be estimated by Eq \ref{eq:carbonsharefuel}.

\begin{equation}\label{eq:carbonsharefuel}
    C_{fu}:C_{in} = \eta \cdot \frac{\varepsilon_{fu}}{\varepsilon_{sb}}
\end{equation}

The rest is assumed to end up as CO$_2$, of which a part $\varepsilon_{s}$ is separated, captured and stored with an efficiency $\eta_{\varepsilon}$, with the remainder $\varepsilon_{v}$ being vented as CO$_2$ to the atmosphere in the exhaust gas.

The biogas produced from digestible biomass is assumed to contain 60 vol-\% CH$_4$ ($e$=50 GJ t$^{-1}$, $\rho=$0.657 \ kg/m$^3_n$) and 40 vol-\% CO$_2$ ($\rho=$1.98 \ kg/m$^3_n$), which calculates to 0.0868	tCO$_2$/MWh$_{CH_4}$. The feedstock input potentials and costs for biogas are given for MWh$_{CH_4}$, and thus MWh$_{in}$ = MWh$_{out}$ for the carbon balance calculations. Thereby the C-content in the slush can be omitted, thus avoiding system boundary issues with the agricultural sector.

The carbon balance equals zero (Eq. \ref{eq:cbalance}):
\begin{equation}\label{eq:cbalance}
    \Delta\varepsilon = \varepsilon_{at} + \varepsilon_s + \varepsilon_v + \varepsilon_{fu} \cdot \eta = 0
\end{equation}

\begin{table}[ht]
    \centering
    \begin{footnotesize}
    \begin{tabular}{l l c c c c c c c c}
    \toprule
         & Type & e & $\eta$ & $\varepsilon_{at}$ & $C_{fu}$ : $C_{in}$ & $\varepsilon_{s}$ & $\varepsilon_{v}$ & e$_{fu}$ & $\varepsilon_{fu}$ \\
         & & GJ/t & & tCO$_2$/MWh$_{in}$ & \% & tCO$_2$/MWh$_{in}$ & tCO$_2$/MWh$_{in}$ & GJ/t & tCO$_2$/MWh \\
         \midrule 
    BioSNG & solid & 18 & 0.7 & -0.3667 &  37.8 & 0.2235 & 0.0046 & 50 & 0.198\\
    BtL & solid & 18 & 0.4 & -0.3667 & 28 & 0.2585 & 0.0053 & 44 & 0.2571\\
    Biogas & dig. & - & 1 & -0.198-0.0868 & 69.5 & 0.085064 & 0.001736 & 50 & 0.198\\
    \bottomrule
    \end{tabular}
    \caption{Carbon balances of bioenergy options. Dig. = digestible.}
    \end{footnotesize}
    \label{tab:carbon_balance}
\end{table}

For the Fischer-Tropsch and methanation processes based on H$_2$ and CO$_2$ inputs, the CO$_2$ is assumed to be cycled within the process, and thus the input-output-ratio of carbon is unity, bar CO$_2$ leakage.

\section{Estimating fuel supply cost}\label{sec:estimateFuelCost}
It is challenging to isolate the liquid fuel (or any other) part of the system cost, since electrofuels are interlinked with the electricity system and both electrofuels and biofuels are connected via carbon capture. Electrolysers are flexible and thus use electricity when it is cheaper.

We provide an estimate for the electricity cost carried by electrolysers as follows. The total electricity cost is the sum of $C_{j,i}^{el}$ for all technologies $j\in(g,tn)$ (electricity generators $g$ and transmission infrastructure $tn$). The total load-weighted electricity price paid by electrolysers (electrolyser $k$ electricity demand $\delta_{i,k,t}^{el}$ multiplied by electricity price $p_{i,t}^{el}$) is divided by the total load-weighted electricity price paid (electricity production  from each generator $j$, multiplied by price $p_{i,t}^{el}$) for each time step $t$ and node $i$. This provides a weight with which the cost of electricity carried by electrolysers is derived.

The share of H$_2$ used for electrofuel production is calculated by dividing the electrofuel H$_2$ demand $\delta_{i,j,t}^{H_2}$ for all electrofuel technologies $j\in F_e$, by the total H$_2$ production $\pi_{i,t}^{H_2}$. The cost of H$_2$ (including electricity, electrolyser capital costs and H$_2$ pipeline costs $C_{j\in H_2}$) is assigned to electrofuels by the share of H$_2$ used for electrofuel production.

The CO$_2$ shadow price $p^{CO_2}$ is used as the cost of CO$_2$ input for electrofuels $\delta_{j}^{CO_2}$ with $j\in F_e$. The total cost calculation is shown in Equation \ref{eq:efuelCost}.

\begin{equation}\label{eq:efuelCost}
    C_{tot}^{F_e} = \sum\limits_{i} C_{i}^{F_e} + \frac{\sum\limits_{i,j\in F_e,t}\delta_{i,j,t}^{H_2}}{\sum\limits_{i,t} \pi_{i,t}^{H_2}} \left( \sum\limits_{j,i}C_{j,i}^{el} \frac{\sum\limits_{i,t} \delta_{i,k,t}^{el} p_{i,t}^{el}}{\sum\limits_{i,t} \pi_{i,j,t}^{el} p_{i,t}^{el}} + \sum_j C_{j\in H_2}\right) + \sum\limits_{j\in F_e}\delta_{j}^{CO_2} p^{CO_2} 
\end{equation}

The cost of solid biomass $b_s$ used to produce biofuels is assigned to the biofuels by the amount of solid biomass used for biofuels $\delta_{j\in F_b,i,t}^{b_s}$, with $j\in F_b$, divided by the total amount of solid biomass used $\pi_{i,t}^{b_s}$. This is added to the capital cost of biomass to liquid $C_{i}^{F_b}$ (Equation \ref{eq:biofuelCost}).

\begin{equation}\label{eq:biofuelCost}
    C_{tot}^{F_b} = \sum\limits_{i} C_{i}^{F_b} + \frac{\sum\limits_{i,j\in F_b,t}\delta_{i,j,t}^{b_s}}{\sum\limits_{i,t} \pi_{i,t}^{b_s}} \sum\limits_{i,t} C_{i,t}^{b_s}
\end{equation}

The CO$_2$ shadow price $p^{CO_2}$ is used as the cost of CO$_2$ emissions for fossil fuels $\varepsilon_{i,j,t}^{CO_2}$ with $j\in F_f$ (fossil fuels), see Equation \ref{eq:emissionCost}, but is is shown separately and not directly assigned to the cost of fossil fuels.

\begin{equation}\label{eq:emissionCost}
    C_{\varepsilon}^{F_f} = \sum\limits_{j\in F_f,i,t}\varepsilon_{i,j,t}^{CO_2} p^{CO_2}
\end{equation}

All of the reassigned costs are subtracted at the appropriate place elsewhere in the system. An opportunity cost is calculated as the total energy system cost increase compared to the base cases, minus the fuel cost difference compared to the base cases. This can sometimes leads to negative opportunity costs, i.e.\ the rest of the energy system is less costly compared to in the base case.

\section{Liquid fuel cost today}\label{sec:fuelCostToday}
The Eurozone weighted average consumer price of Euro-super 95 (gasoline) in 2019 was 0.56 \texteuro/l excluding taxes and levies, 0.61 \texteuro/l for diesel and 0.44 \texteuro/l for low-sulphur fuel oil \cite{EU2021}. The average price of jet fuel was 0.44 \texteuro/l \cite{IndexMundi2022}, and the share of diesel in road transport was around 70\% in 2018 \cite{FuelsEurope2018}. The total cost of \textit{transport fuels} in 2018 is estimated at 282 billion~\texteuro, based on the data summarized in Table \ref{tab:fuel_costs}.

\begin{table}[ht]
    \centering
    \begin{footnotesize}
    \begin{tabular}{lcccc}
        \toprule
         & Cost & Energy density & Demand 2018 & Total cost\\ 
         & \texteuro/l & MJ/l & TWh & billion~\texteuro\\ 
         \midrule 
        Gasoline & 0.56 & 33 & 1083 & 61\\
        Diesel & 0.61 & 36 & 2528 & 168\\
        Fuel oil & 0.44 & 39 & 629 & 26\\
        Jet fuel & 0.44 & 35 & 612 & 28\\
        \cmidrule{5-5} \relax
         & & & & $\Sigma$ \ 282 \\ 
        \bottomrule 
    \end{tabular}
    \caption{Costs and demands of different transport fuels in 2018}
    \label{tab:fuel_costs}
    \end{footnotesize}
\end{table}

\clearpage

\bibliographystyle{elsarticle-num}

\biboptions{sort&compress}

\end{document}